\documentclass[12pt,a4paper]{article}
\pdfoutput=1
\usepackage{jheppub,psfrag,slashed,cancel,lscape,caption,array,graphicx,subcaption,todonotes,tikz,fp}
\usepackage[utf8]{inputenc}
\usepackage{amsmath}
\usepackage{mathtools}
\usepackage{mathrsfs}
\usepackage{multirow}
\usepackage{booktabs} %nice tables
\usepackage[normalem]{ulem}

\usetikzlibrary{decorations.pathmorphing,positioning,snakes,calc,fadings}
\usetikzlibrary{decorations.markings}

\DeclareFontFamily{OT1}{pzc}{}
\DeclareFontShape{OT1}{pzc}{m}{it}{<-> s * [1.350] pzcmi7t}{}
\DeclareMathAlphabet{\mathpzc}{OT1}{pzc}{m}{it}

\definecolor{myblue}{rgb}{0,0,0.7}

\definecolor{mygreen}{rgb}{0,0.4,0}

\definecolor{myred}{rgb}{0.4,0,0}

\def\cv{\alpha}

\def\cL{\mathcal{L}}
\def\cN{\mathcal{N}}
\def\cO{\mathcal{O}}
\def\cD{\mathcal{D}}

\def\cV{\mathcal{V}}
\def\cH{\mathcal{H}}
\def\cF{\mathcal{F}}

\def\cA{\mathcal{A}}

\def\eps{\epsilon}
\def\bC{\bar{C}}
\def\bP{\bar{P}}

\def\cAL{\cA_{\rm L}}
\def\cAR{\cA_{\rm R}}

\def\nn{\nonumber}

\def\eqn#1{eq.~(\ref{#1})}

\def\sec#1{section~{\ref{#1}}}

\def\braket#1{\langle #1 \rangle}

\DeclareMathOperator{\Tr}{Tr}

\preprint{UUITP-24/18 \\
\phantom{~} \hfill NORDITA 2018-042}

\title{Unraveling conformal gravity amplitudes}

\author[a,b]{Henrik Johansson,}
\author[a]{Gustav Mogull}
\author[a]{and Fei Teng}
\affiliation[a]{Department of Physics and Astronomy, Uppsala University, 75108 Uppsala, Sweden}
\affiliation[b]{Nordita, Stockholm University and KTH Royal Institute of Technology,\\ Roslagstullsbacken 23, 10691 Stockholm, Sweden}
\emailAdd{henrik.johansson@physics.uu.se}
\emailAdd{gustav.mogull@physics.uu.se}
\emailAdd{fei.teng@physics.uu.se}

\abstract{Conformal supergravity amplitudes are obtained from the double-copy construction using gauge-theory amplitudes, and compared to direct calculations starting from conformal supergravity Lagrangians. We consider several different theories: minimal ${\cal N}=4$ conformal supergravity, non-minimal ${\cal N}=4$ Berkovits-Witten conformal supergravity, mass-deformed versions of these theories, as well as supersymmetry truncations thereof. Coupling the theories to a Yang-Mills sector is also considered. For all cases we give the gravity Lagrangians that the double copy implicitly generates. The two main results are: we determine a Lagrangian for the non-minimal Berkovits-Witten theory, and we uncover the double-copy prescription for the minimal ${\cal N}=4$ conformal supergravity.}


\begin{document}

\tikzset{
  boson/.style={
    decoration={snake, segment length=2mm, amplitude=0.5mm},
    decorate,
  },
  directedScalar/.style={
    dashed,
    postaction={decorate},
    decoration={markings,mark=at position 0.5 with {\arrow[scale=1.5]{>}}}
  },
  gluon/.style={decorate,
    decoration={coil,amplitude=5pt, segment length=5.5pt,  pre length=.1cm, post length=.1cm}
  },
  scalar/.style={
    dashed
  }
}

\maketitle
\flushbottom

%**************************************************
\section{Introduction}

Quantum field theories with four-derivative kinetic terms are notorious for their enigmatic behavior. On the one hand they generically contain ghost-like states associated to indefinite signs in the propagators, suggesting that they are non-unitary theories that should be discarded from further study. On the other hand they may have exceptionally good behavior in the ultraviolet regime, which over the years has stimulated attempts to solve ultraviolet problems in the standard framework of two-derivative theories that currently describe Nature. 

Well-known classes of interesting four-derivative models are (super)gravity theories built out of Riemann curvature squared invariants, $R^2$, which are examples of renormalizable theories of gravity~\cite{Stelle:1976gc}.  Conformal (super)gravity~\cite{Weyl, Kaku:1977pa, Ferrara:1977ij, Kaku:1978nz, Bergshoeff:1980is} takes a special place among these theories, due to its high degree of symmetry which enlarges the local Poincar\'e symmetry of Einstein gravity to local conformal (Weyl) symmetry. Gravities of $R^2$ type also feature in cosmology, with the Starobinsky model of inflation as the prime example~\cite{Starobinsky:1980te}.  Other interesting four-derivative models are gauge theories of Lee-Wick type~\cite{Lee:1969fy,Lee:1970iw}, which are extensions to the Pauli-Villars regularization trick and which have been employed in attempts to solve the hierarchy problem of the Standard Model~\cite{Grinstein:2007mp}.

In this paper we study tree-level scattering amplitudes of several four-derivative models, with a focus on conformal (super)gravity and closely-related four-derivative gauge theories. Previous work~\cite{Johansson:2017srf} has shown that scattering amplitudes in conformal (super)gravity, of the non-minimal Berkovits-Witten type~\cite{Berkovits:2004jj}, can be obtained from the Bern-Carrasco-Johansson (BCJ) double copy of two gauge theories~\cite{Bern:2008qj,Bern:2010ue}. The two gauge theories are (super-)Yang-Mills theory and a new gauge theory with a four-derivative kinetic term of the form $(DF)^2$~\cite{Johansson:2017srf}. The double-copy construction offers better means for understanding conformal gravity at both the classical and quantum level, as it maps a complex gravitational theory to an easier-to-study gauge theory. 

The double copy is currently best understood as a consequence of a duality between color and kinematics~\cite{Bern:2008qj,Bern:2010ue}, which is a property of a large variety of different gauge theories~\cite{Bern:2008qj,Bern:2010ue,Johansson:2014zca,Johansson:2015oia,Bargheer:2012gv, Chiodaroli:2014xia, Chiodaroli:2015rdg, Johansson:2017srf}. Given that two gauge theories obey the duality, the color factors in the amplitudes of the first theory can be replaced by the kinematic numerator factors of the second theory~\cite{Bern:2008qj,Bern:2010ue}. Doing so ``doubles up'' the spin of the particles, and promotes the gauge invariance to a diffeomorphism invariance~\cite{Chiodaroli:2017ngp}, thus giving amplitudes that describe the scattering of spin $\le2$ states in a gravitational theory. By now, vast classes of gravitational theories are understood from the double-copy perspective~\cite{Bern:2010ue,Bern:2011rj,Bern:2012uf,Carrasco:2012ca,Broedel:2012rc,Bargheer:2012gv,Huang:2012wr,Chiodaroli:2013upa,Anastasiou:2013hba,Johansson:2014zca,Chiodaroli:2014xia,Chiodaroli:2015rdg,Chiodaroli:2015wal,Anastasiou:2016csv,Chiodaroli:2017ngp,Anastasiou:2017nsz,Johansson:2017bfl, Johansson:2017srf,Chiodaroli:2017ehv,Azevedo:2017lkz,Azevedo:2018dgo}.

The duality provides a rich structure to tree-level amplitudes~\cite{Mafra:2011kj,Broedel:2012rc,Huang:2013kca, Cachazo:2012uq,delaCruz:2015dpa,Mafra:2015vca,Bjerrum-Bohr:2016axv,Chiodaroli:2017ngp,Du:2016wkt,Du:2017kpo,Du:2017gnh,Teng:2017tbo, Plefka:2018zwm, Chen:2017bug}, most notably through the color-ordered $n$-point gluon amplitudes, which are constrained by the so-called BCJ relations~\cite{Bern:2008qj,BjerrumBohr:2009rd,Stieberger:2009hq} --- these can be used to eliminate all but $(n-3)!$ independent amplitudes.  At the quantum level, the duality interrelates the kinematic numerators of various loop diagrams, making it possible to obtain most of them in terms of a small number of master diagrams~\cite{Bern:2010ue,Carrasco:2011mn,Bern:2012uf,Carrasco:2012ca,Bjerrum-Bohr:2013iza,Bern:2013yya,Nohle:2013bfa,Chiodaroli:2013upa,Johansson:2014zca,Mafra:2015mja,Johansson:2015oia,He:2015wgf,Mogull:2015adi, Johansson:2017bfl}. When the double copy is applied to gauge-theory tree amplitudes with external adjoint particles it becomes equivalent to the well-known Kawai-Lewellen-Tye (KLT) formula~\cite{Kawai:1985xq,Bern:1998sv,BjerrumBohr:2010hn}. Whereas for non-adjoint and loop-level amplitudes the double copy provides a more general framework, which has lead to rapid advances in gravitational loop-level calculations~\cite{Bern:2010ue,Carrasco:2011mn,Bern:2011rj,BoucherVeronneau:2011qv,Boels:2011tp,Bern:2012uf,Bern:2012uc,Carrasco:2012ca,Boels:2013bi,Bjerrum-Bohr:2013iza,Bern:2013yya,Nohle:2013bfa,Chiodaroli:2013upa,Johansson:2014zca,Chiodaroli:2014xia,Mafra:2015mja,He:2015wgf,Mogull:2015adi, Chiodaroli:2015rdg, Luna:2016idw,He:2016mzd,He:2017spx,Johansson:2017bfl, Nandan:2018ody} and related ultraviolet studies~\cite{Bern:2012uf,Bern:2012gh,Bern:2013uka,Bern:2013qca,Bern:2014lha, Bern:2017yxu,Bern:2017ucb,Bern:2018jmv}.

Color-kinematics duality and the double copy have found recent applications to off-shell structures and quantities, such as form factors~\cite{Boels:2012ew, Boels:2015yna,Yang:2016ear,Boels:2017skl}, classical solutions~\cite{Saotome:2012vy,Monteiro:2014cda,Luna:2015paa,Ridgway:2015fdl,White:2016jzc,Luna:2016due,Cardoso:2016amd,Luna:2016hge,Goldberger:2016iau,Goldberger:2017frp,Adamo:2017nia,Goldberger:2017vcg, Goldberger:2017ogt, Bahjat-Abbas:2017htu, Carrillo-Gonzalez:2017iyj, Li:2018qap}, symmetries~\cite{Borsten:2013bp,Anastasiou:2013cya,Anastasiou:2013hba,Anastasiou:2014qba,Anastasiou:2015vba,Chiodaroli:2016jqw,Chiodaroli:2017ngp,Arkani-Hamed:2016rak, Ferrara:2018iko} and the kinematic Lie algebra~\cite{Bern:2010yg,Monteiro:2011pc,Cheung:2016prv}. The duality and double copy also prominently feature in amplitudes of string theory~\cite{Stieberger:2009hq,BjerrumBohr:2010hn, Mafra:2011nw,Mafra:2012kh,Broedel:2013tta,Huang:2016tag, Carrasco:2016ldy,Carrasco:2016ygv, Tourkine:2016bak,Hohenegger:2017kqy, Azevedo:2018dgo,Geyer:2018xwu}, the non-linear-sigma model, Born-Infeld, Volkov-Akulov, the special galileon theory~\cite{Chen:2013fya,Cheung:2014dqa,Cachazo:2014xea,Cachazo:2016njl,Du:2016tbc,Cheung:2016prv,Carrasco:2016ldy, Cheung:2017yef}, and higher-spin theory~\cite{Ponomarev:2017nrr}.

In this paper we are ultimately interested in the properties of ${\cal N}=4$ conformal supergravity, which has been argued to have the maximal degree of supersymmetry compatible with four-dimensional local conformal symmetry~\cite{deWit:1978pd}. Witten's twistor string theory~\cite{Witten:2003nn} --- which gives tree amplitudes of ${{\cal N}=4}$ super-Yang-Mills theory in its single-trace sector~\cite{Roiban:2004yf}  --- is well-known to have a multitrace sector that is ``contaminated'' by ${{\cal N}=4}$ conformal supergravity. This non-minimal form of conformal supergravity was first studied in isolation by Berkovits and Witten in ref.~\cite{Berkovits:2004jj}; however, no complete Lagrangian was given there. Recently Tseytlin considered general non-minimal ${{\cal N}=4}$ conformal supergravities, and concluded that they are free of conformal anomalies given that four vector multiplets are added to the spectrum~\cite{Tseytlin:2017qfd}; thus agreeing with the analysis of the minimal theory~\cite{Fradkin:1983tg,Fradkin:1985am,Romer:1985yg}. A complete bosonic action for all ${\cal N}=4$ non-minimal conformal supergravity theories --- parametrized by a free function --- has been proposed by Butter, Ciceri, de Wit and Sahoo~\cite{Butter:2016mtk}. We confirm in this work, by direct calculation of amplitudes from both the double copy and the action, that the Lagrangian of the Berkovits-Witten theory is given by a simple choice of the free function. 

Somewhat surprisingly, the minimal version of ${{\cal N}=4}$ conformal supergravity has a trivial tree-level S-matrix when restricted to physical planewave states in four-dimensional flat space~\cite{Maldacena:2011mk,Adamo:2013tja,Adamo:2016ple,Beccaria:2016syk}. While this is a highly interesting behavior, perhaps offering some mitigation to the unitarity problem, it is an obstruction if we wish to better understand the theory by studying its scattering amplitudes. In particular, to determine whether minimal ${{\cal N}=4}$ conformal supergravity is constructible from a double-copy perspective one needs non-trivial amplitudes for comparisons.  This problem can be circumvented by considering loop amplitudes for physical planewave states, or tree-level amplitudes for non-planewave states. We consider the latter case, even if such non-planewave states have been suggested to be problematic due to their growing behavior at infinity~\cite{Adamo:2018srx}. Inspecting the non-vanishing amplitudes, we confirm that there exists a double-copy construction for minimal ${{\cal N}=4}$ conformal supergravity; it involves a minimal version of the $(DF)^2$ gauge theory where only the kinetic term is retained. 

As expected, the minimal $(DF)^2$ theory also has a trivial tree-level S-matrix for planewave states~\cite{Johansson:2017srf}, which we explain in terms of the classical field equations along the lines of Maldacena's argument~\cite{Maldacena:2011mk}. Inspired by this, we note that ``minimal'' four-derivative theories in general can be constructed to have a trivial tree-level S-matrix for planewave states. We illustrate this property through a four-derivative scalar toy model. Interestingly, the minimal scalar, gauge and gravity theories all can be {\it mass deformed} by adding two-derivative theories to their Lagrangians, corresponding to $\phi^3$, Yang-Mills and Einstein theories, respectively. The four-derivative tree-level planewave S-matrices then become identified with the corresponding two-derivative S-matrices, up to the overall mass scale that parametrizes the deformation. This is analogous to the mechanism in Anti-de-Sitter space described by Maldacena~\cite{Maldacena:2011mk}.

All results obtained here for the ${{\cal N}=4}$ conformal supergravity theories also apply to supersymmetry truncations of these theories. In particular, corresponding ${{\cal N}=0,1,2}$ conformal (super)gravity theories are known to exist \cite{Townsend:1979ki,Kaku:1978nz,Bergshoeff:1980is, Kaku:1977pa,Ferrara:1977mv}, and we obtain them as double copies by attributing the supersymmetry to the ${{\cal N}=0,1,2}$ (super-)Yang-Mills side of the double copy.  The other side of the double copy is the bosonic $(DF)^2$ theory~\cite{Johansson:2017srf}, which comes in various forms, as summarized in the following table of double copies considered in this paper: 
\begin{center}
\begin{tabular}{c|c|c|c}
\text{double copy}& $m\rightarrow 0$ & finite $m$ & $m\rightarrow\infty$ \\ \hline
  $\big((DF)_{\text{min.}}^2+\text{YM}\big)$ $\otimes$ SYM & min.\,CG   &  min.\,Weyl-Einstein  & Einstein \\
  $\big((DF)^2+\text{YM} \big)$ $\otimes$ SYM &  CG & Weyl-Einstein & Einstein \\
  $\big((DF)^2+\text{YM}+\phi^3\big)$ $\otimes$ SYM & Weyl-YM & Weyl-Einstein-YM & Einstein-YM
\end{tabular}
\end{center}
The mass parameter interpolates between two-derivative and four-derivative theories, and the inclusion of self-interacting scalars in the bosonic gauge theories translates to the inclusion of non-abelian Yang-Mills (YM) sectors in the gravity theories~\cite{Chiodaroli:2014xia,Johansson:2017srf} (see also ref.~\cite{Azevedo:2018dgo} for similar string theory double copies). Weyl-Yang-Mills conformal supergravities were first described in refs.~\cite{Ferrara:1977ij,Kaku:1977rk,Das:1978nr}. 

This paper is organized as follows: In \sec{sec:toymodel}, we discuss technical details of scalar four-derivative theories as a warmup to the more interesting gauge and gravitational theories.  In \sec{sec:review}, we review the conformal-gravity double-copy construction of ref.~\cite{Johansson:2017srf}, and in \sec{sec:einsteingravity} we consider details of ${\cal N}=4$ Einstein supergravity that we need for later purposes. The double-copy construction of minimal conformal supergravity is given in~\sec{sec:mintheories}, and in~\sec{sec:confgravity} we compute amplitudes directly from a non-minimal conformal supergravity Lagrangian and determine the precise details of the Berkovits-Witten theory.

%**************************************************
\section{Warmup: four-derivative scalar theories}
\label{sec:toymodel}

We here discuss a four-derivative scalar toy model that illustrates some of the salient features that we will encounter when dealing with higher-derivative gauge theories and supergravities.  After discussing linearized on-shell solutions, propagators and formal aspects of scattering in generic four-derivative scalar theories, we specialize to a specific toy model that at low energy behaves as $\phi^3$ theory, and which is marginal in $D=6$.  At high energies, it has a superficial behavior consistent with a theory marginal in $D=8$.  This theory is carefully constructed (tuned) so as to make it a prototype for minimal conformal supergravity coupled to Einstein supergravity.

\subsection{On-shell states}\label{sec:onshellstates}

Consider a massless scalar field with a four-derivative kinetic term $\cL=-\frac12\phi\square^2\phi+\ldots$.  The linearized four-derivative equation of motion, $\square^2\phi=0$, has two independent solutions parametrized by an on-shell momentum $p^2=0$:
\begin{align}\label{2sol}
\phi_{\rm pw}(x)=e^{ip\cdot x}\,,& &\phi_{\cancel{\rm pw}}(x) = i \left(\cv\cdot x\right) e^{ip\cdot x}\,,
\end{align}
where $\cv^\mu$ is a constant vector satisfying $\cv\cdot p\neq0$.\footnote{It may appear that the freedom in $\cv^\mu$ corresponds to a larger family of solutions; however, note that the converse constraint $\tilde{\cv} \cdot p =0$ defines a $(D-1)$-dimensional space of vectors $\tilde{\cv}^\mu_i$, $i=1,\ldots, D-1$.  Hence $\cv \cdot p \neq 0$ effectively defines a one-dimensional space orthogonal to $\tilde{\cv}^\mu_i$. As an alternative perspective, note that $\phi_{\cancel{\rm pw}}$ with a different constant vector $\cv'$ corresponds to a superposition of the two solutions in \eqn{2sol}.}  The first solution, which satisfies $\square\phi_\text{pw}=0$, is the usual planewave mode that is also present in two-derivative scalar theories.  The second, for which $\square\phi_{\cancel{\rm pw}}\neq0$, is a non-planewave mode that is specific to four-derivative theories.\footnote{Note that $\varphi\equiv\Box \phi_{\cancel{\rm pw}}$ is a planewave since $\Box\varphi=0$.}  As this mode grows linearly in $x$, it is questionable whether it can acceptably be taken as an external state of the S-matrix, since asymptotic states are taken to live at infinity.  Also, the growing behavior of the $\phi_{\cancel{\rm pw}}$ mode prevents orthogonalization of the two states~(\ref{2sol}). See refs.~\cite{Adamo:2018srx, Farrow:2018yqf} for recent work related to non-planewave modes. 

Related to the orthogonalization problem, the off-shell propagator in this theory,
\begin{equation}\label{eq:p4prop}
\begin{tikzpicture}
  [baseline={([yshift=-.5ex]current bounding box.center)},thick,inner sep=0pt,minimum size=0pt,>=stealth,scale=0.5]
  \node (1) at (0,0) {};
  \node (2) at (4,0) {};
  \draw[scalar] (1) node[left=0.1] {$\phi$} to node[below=0.2] {$p$} (2) node[right=0.1] {$\phi$} {};
\end{tikzpicture}
=-\frac{i}{p^4}\,,
\end{equation}
consists of only one term, a double pole, that does not obviously distinguish between the planewave and non-planewave modes.  To resolve the double pole, instead consider mass deforming the Lagrangian to $\cL=-\frac12\phi\square(\square+m^2)\phi+\ldots$, with the eventual intention of sending the mass parameter $m\to0$.  The two solutions to the linearized equation of motion, $\square(\square+m^2)\phi=0$, are a massless and a massive planewave:
\begin{align}
\begin{aligned}\label{2pw}
\phi_0(x) &= e^{i p\cdot x}\,,  & p^2&=0\,,\\
\phi_m(x) &= e^{i p_m\cdot x}\,, &  p_m^2&=m^2\,,
\end{aligned}
\end{align}
which satisfy $\Box \phi_0 =0$ and $(\Box+m^2) \phi_m=0$.  The mass-deformed propagator is 
\begin{align}
\begin{tikzpicture}
  [baseline={([yshift=-.5ex]current bounding box.center)},thick,inner sep=0pt,minimum size=0pt,>=stealth,scale=0.5]
  \node (1) at (0,0) {};
  \node (2) at (4,0) {};
  \draw[scalar] (1) node[left=0.1] {$\phi$} to node[below=0.2] {$p$} (2) node[right=0.1] {$\phi$} {};
\end{tikzpicture}
= -\frac{i}{p^2(p^2-m^2)} =  \frac{1}{m^2} \left(  \frac{i}{p^2} -\frac{i}{p^2-m^2} \right)\,,
\label{mprop}
\end{align}
hence the two poles are well-separated after partial fractioning.

The one-to-one match between states and propagator terms makes it straightforward to identify the states in scattering amplitudes by examining the types of poles that appear.  However, the relative sign implies that the massive mode is ghostlike (assuming that the massless mode is physical).  Indeed, as is well known, the four-derivative kinetic term suggest that we are dealing with a non-unitary theory.

In the strict $m\to0$ limit the two linearized solutions (\ref{2pw}) become identical, so one needs to consider subleading terms in the $m\to0$ limit in order to still have two modes.  We expand the massive wavefunction $\phi_m$ around $m=0$ using $p^\mu_m = p^\mu +\frac{m^2}{2 p\cdot q} q^\mu$, where $p^\mu$ and $q^\mu$ are null vectors independent of $m$ with $p\cdot q\neq0$:
\begin{equation}
\phi_m(x) = e^{ip_m\cdot x} =
e^{ip\cdot x}\,\Big(1+\frac{i m^2}{2 p\cdot q} q\cdot x + {\cal O} (m^4)  \Big)\,.
\end{equation}
The non-planewave mode $\phi_{\cancel{\rm pw}}$ emerges in the $m\to0$ limit as the linear combination
\begin{equation}\label{eq:npwLimiting}
\phi_{\cancel{\rm pw}}(x)=\frac{\phi_m(x) - \phi_0(x) }{m^2}\Big|_{m^2\rightarrow 0} =
i (\cv \cdot x)  e^{ip\cdot x}\,,
\end{equation}
where we have the identification $\cv^\mu=\frac{q^\mu}{2 p\cdot q}$.  From this formula it is also clear that we can view the non-planewave state as the $m^2$ derivative of the massive planewave state in the neighborhood of $m^2=0$. Alternatively, we can view the non-planewave state as the momentum derivative of the massless plane wave $\phi_{\cancel{\rm pw}}=(\cv\cdot\partial_p)\phi_{\rm pw}$, which is a more conventional interpretation~\cite{Berkovits:2004jj,Adamo:2016ple,Farrow:2018yqf}.  

\subsection{Scattering amplitudes from classical solutions}\label{sec:scalaramp}

A well-known aspect of tree-level scattering amplitudes is that they can be read out from a perturbative solution to the classical equations of motion~\cite{Boulware:1968zz,Monteiro:2011pc}, a procedure which was streamlined in gauge theories by the Berends-Giele recursion~\cite{Berends:1987me}.  In a two-derivative scalar theory (for example, $\phi^3$ theory) the equation of motion is solved order-by-order in momentum space:
\begin{align}
\phi(p)=\sum_{n=0}^\infty\phi^{(n)}(p)\,,
\end{align}
where $\phi^{(n)}(p)$ is of $n$th order in the coupling constant.  The zeroth-order solution $\phi^{(0)}(p)$, which solves the free equation of motion $\square\phi^{(0)}=0$, has support only on $p^2=0$ (hence $\phi^{(0)}$ is a planewave).  Using functional differentiation of the higher-order solutions one obtains an off-shell Berends-Giele current
\begin{align}\label{eq:funcDeriv}
J(p_1,p_2,\ldots,p_n)\equiv
\frac{\delta^{n-1}\phi^{(n-2)}(-p_1)}{\delta \phi^{(0)}(p_2)\delta \phi^{(0)}(p_3)\ldots\delta \phi^{(0)}(p_{n})}\,.
\end{align}
An $n$-point tree-level amplitude is then obtained using the LSZ prescription by multiplying with an inverse propagator  corresponding to the off-shell leg 1,
\begin{align}
\cA^{(2)}_n(1,2,\ldots, n)=-i\lim_{p_1^2\rightarrow 0}p_1^2 \, J(p_1,p_2\ldots,p_n)\,,
\end{align}
where the superscript (2) is used to indicate a two-derivative theory.

In a four-derivative mass-deformed scalar theory the details are mostly the same.  However, with the free equation of motion being $\square(\square+ m^2)\phi^{(0)}=0$, the zeroth-order solution $\phi^{(0)}$ can have support on either $p^2=0$ or $p^2=m^2$ for each external leg, corresponding to either massless or massive planewaves being scattered.  As we in the current work do not seek to scatter more than one ghostlike state we can make the simplifying assumption that massless planewaves are used as boundary conditions.

For the off-shell leg 1 there are similarly two choices available,
\begin{subequations}\label{eq:BGdeformed}
\begin{align}
  \label{eq:BGdeformeda}
  \mathcal{A}^{(4)}_n(1,2,\ldots, n)&=i\lim_{p_1^2\rightarrow 0}p_1^2(p_1^2-m^2) \, J(p_1,p_2,\ldots,p_n)\,,\\
  \label{eq:BGdeformedb}
  \mathcal{A}^{(4)}_n(1_m,2,\ldots, n)&=i\!\!\lim_{p_1^2\rightarrow m^2}p_1^2(p_1^2-m^2) \, J(p_1,p_2,\ldots,p_n)\,,
\end{align}
\end{subequations}
where the superscript (4) reminds us that we are dealing with a four-derivative theory. The first possibility, which identifies leg $p_1$ as a massless planewave, is easily seen to be equivalent to a multiplication with the inverse massless propagator $-i p_1^2 m^2$, matching the corresponding term in the partial-fractioned propagator~(\ref{mprop}).  The second possibility, which gives a massive planewave, is equivalent to multiplying by $i (p_1^2-m^2) m^2$, which is the inverse massive propagator.  

To convert the above amplitudes into one involving a non-planewave mode, one repeats the limiting procedure given  for the linearized states in eq.~(\ref{eq:npwLimiting}), obtaining
\begin{equation}\label{eq:onenpw}
\cA_n^{(4)}(\tilde{1},2,\ldots, n)=
\lim_{m^2\rightarrow 0}\frac{\cA^{(4)}_n(1_m,2,\ldots, n)-\cA^{(4)}_n(1,2,\ldots, n)}{m^2}\,.
\end{equation}
One might be tempted to think of this as an $m^2$ derivative of the massive amplitude; however, this is not the case because the various limits taken in defining the amplitudes do not commute.

Consider the amputated current $\widehat J(p_1,p_2,\ldots,p_n)\equiv i p_1^2(p_1^2-m^2) \, J(p_1,p_2,\ldots,p_n)$, which schematically has the form
\begin{equation}
\widehat J(p_1,p_2,\ldots,p_n) = A + B (p_1^2-m^2) + C p_1^2 + \ldots\,,
\label{schematic1}
\end{equation}
where the suppressed terms are higher powers of $p_1^2$ and $(p_1^2-m^2)$; $A,B,C$ are functions independent of $p_1^2$ and $m^2$. The massless amplitude is obtained by setting $p_1^2=0$, and the massive one from $p_1^2=m^2$; thus, they have the expansions
\begin{align}
\cA^{(4)}_n(1,2,\ldots, n) & =  A - m^2 B +{\cal O}(m^4)\,, \nn \\
\cA^{(4)}_n(1_m,2,\ldots, n) &= A+ m^2 C+{\cal O}(m^4)\,.
\label{schematic2}
\end{align}
In the $m^2 \rightarrow 0$ limit the term $A$ is the planewave amplitude, and following \eqn{eq:onenpw} the sum $B+C$ is the non-planewave amplitude, which is not the $m^2$ derivative of either amplitude in \eqn{schematic2}. However, from inspecting \eqn{schematic1},  it is clear that we can obtain $B+C$ from a $p_1^2$ derivative directly on the amputated current:
\begin{equation}
\cA_n^{(4)}(\tilde{1},2,\ldots, n) = \lim_{p_1^2\rightarrow 0}   \frac{\partial}{\partial p_1^2} \widehat J(p_1,p_2,\ldots,p_n) = \lim_{p_1^2 \rightarrow 0}  \cv^\mu \frac{\partial}{\partial p_1^\mu} \widehat J(p_1,p_2,\ldots,p_n)\,, 
\end{equation}
where in the last step we used the chain rule on $p^\mu_{1}|_\text{off-shell}= p^\mu_1|_\text{on-shell} + p^2_1 \, \cv^\mu$ to express the derivative directly in terms of the momentum, which is more convenient than the squared momentum. Note, however,  that we cannot take the limit $p_1^2\rightarrow 0$ before the derivative since then overall factors of $p^2_1$ are undetectable. 

With a few exceptions, in this paper we will be concerned with scattering only massless, physical external states.  The massless planewave modes $\phi_0$ do not depend on $m$; similarly, it does not matter in which order one takes the $m\to0$ and $p^2\to0$ limits in the LSZ procedure.  One may therefore avoid introducing $m$ to the theory entirely and simply use
\begin{equation}\label{eq:4derphysical}
\cA^{(4)}_n(1,2,\ldots, n)=i\lim_{p_1^2\rightarrow 0}p_1^4\, J(p_1,p_2,\ldots,p_n)\,.
\end{equation}
Namely, one can calculate physical amplitudes involving planewave states using the usual LSZ procedure by isolating $1/p^4$ poles on external legs.  In conformal supergravities the mass deformation breaks dilatation symmetry and the $m\to0$ limit is potentially problematic.  We will return to this point in section~\ref{sec:weyl2}.

\subsection{Factorization of four-derivative trees}

To better understand the correspondence between states and propagators in generic four-derivative scalar theories it is helpful to see how their tree amplitudes factorize.  Consider an $n$-point amplitude of massless planewave states in the mass-deformed theory, where $P=p_1+p_2 +\cdots +p_k$ is an internal channel.  Using the Feynman rules, the amplitude factorizes in the $P^2\rightarrow 0$ and $P^2\rightarrow m^2$ limits as
\begin{equation}
\cA^{(4)}_n(1,2,\ldots, n) = \widehat{J}(-P,p_1,p_2, \ldots, p_k)  \frac{-i}{P^2(P^2-m^2)} \widehat{J}(P,p_{k+1}, \ldots, p_n) + {\rm finite}\,,
\end{equation}
where we use amputated currents $\widehat{J}(P,\cdots)\equiv iP^2(P^2-m^2)J(P,\cdots)$.  This decomposes into two contributions using partial fractions,
\begin{align}
\cA^{(4)}_n(1,\ldots, n)=
\cAL(0)\frac{i}{m^2P^2}\cAR(0)+\cAL(m^2)\frac{-i}{m^2(P^2-m^2)}\cAR(m^2)
+{\rm finite},
\label{fact0}
\end{align}
where we use shorthand notation for the amplitudes $\cA_{\rm L}(P^2) \equiv \cA^{(4)}_n(-P ,p_1, \ldots, p_k)$ and $\cA_{\rm R}(P^2) \equiv \cA^{(4)}_n(P,p_{k+1}, \ldots, p_n)$ when $P^2=0$ or $P^2=m^2$.  That the amputated currents $\widehat{J}$ can be replaced with amplitudes follows from the definition~(\ref{eq:BGdeformed}) --- expanding each current around $P^2=0$ or $P^2=m^2$, the leading piece is an amplitude and subleading pieces contribute only to the finite terms.

We are prevented from taking $m\rightarrow0$ limit by the explicit appearance of $m^2$ in the denominators.  However, it is a simple exercise to show that the expression (\ref{fact0}) is precisely equal to
 \begin{align}
\cA^{(4)}_n(1,\ldots, n)&=
\cAL(0)\frac{-i}{P^2 (P^2-m^2)}\cAR(0)+
\cAL(0)\frac{-i}{(P^2-m^2)}\frac{\cAR(m^2)-\cAR(0)}{m^2}\nn\\
&\qquad+\frac{\cAL(m^2)-\cAL(0)}{m^2}\frac{-i}{(P^2-m^2)}\cAR(0)\\
&\qquad+\frac{\cAL(m^2)-\cAL(0)}{m^2}\frac{-im^2}{(P^2-m^2)}
\frac{\cAR(m^2)-\cAR(0)}{m^2} + {\rm finite}\,.\nn
\end{align}
The $m^2$ factors in the denominators are now matched with corresponding vanishing expressions in the numerators, so there are no longer any singularities as $m\rightarrow 0$.  Only the first three terms survive in this limit, and using the definition of amplitudes with a single non-planewave state given in eq.~(\ref{eq:onenpw}) we find that
 \begin{equation}
\cA^{(4)}_n(1,2,\ldots, n)=\cAL(0)\frac{-i}{P^4}\cAR(0)+\cAL(0)\frac{-i}{P^2}\tilde{\cAR}(0)+\tilde{\cAL}(0)\frac{-i}{P^2}\cAR(0)+{\rm finite}\,,
\label{PWfact}
\end{equation}
where the amplitudes carrying tildes are those with non-planewave states in the first slot.  Given that the above factorization is also implied by simple dimensional analysis, we expect it also to hold for more general external non-planewave states. 

From this exercise we learn a number of important lessons about amplitudes in four-derivative massless scalar theories:
\begin{enumerate}
\item Leading factorization poles $1/P^4$ always imply an exchange of planewave states.  
\item Planewave and non-planewave states are mixed together in the subleading factorization poles $1/P^2$, which agrees with our earlier observation that the states cannot be orthogonalized.
\item Subleading poles $1/P^2$ generally have non-unique residues. This is because the non-planewave amplitudes depend on $\cv^\mu$ through the parametrization of the momenta in the neighborhood of the pole,  $P^\mu|_\text{off-shell}=P^\mu|_\text{on-shell}+ P^2\cv^\mu$, which is inherent in the definition of non-planewave states~(\ref{eq:npwLimiting}).
\item If a leading factorization pole $1/P^4$ is absent from an amplitude, the interpretation of states and poles $1/P^2$ in this channel should become the usual one for two-derivative theories.
\end{enumerate}
This last point requires further explanation.  In the factorization~(\ref{PWfact}) if either $\cA_{\rm L}(0)$ or $\cA_{\rm R}(0)$ vanishes as $m \rightarrow 0$ then the $1/P^4$ pole is of course absent; however, subleading $1/{P^2}$ poles might still be present. For example, if one of the amplitudes is proportional to $m^2$ (and therefore vanishes as $m\to0$) then the non-planewave state typically gives a non-vanishing $1/P^2$ pole.  This phenomenon occurs in minimal conformal supergravities, and we will now elaborate on it using an explicit toy model.

\subsection{A four-derivative toy model}\label{sec:ttoymodel}

To illustrate some of the features discussed,  we now consider a specific four-derivative toy model with Lagrangian
\begin{equation}\label{Lphi3deformed}
\cL^{(4)}=-\frac{1}{2}\left(\square\phi\right)^2+\frac{g}{2}\phi^2\square\phi-
\frac{g^2}{8}\phi^4+m^2\left(\frac{1}{2}(\partial_\mu\phi)^2+\frac{g}{3!}\phi^3\right)\,,
\end{equation}
where $g$ is the coupling constant. As is apparent, the four-derivative theory includes a mass deformation by the ordinary two-derivative $\phi^3$ theory: $\cL^{(2)}=\frac12(\partial_\mu\phi)^2+\frac g{3!}\phi^3 $.  The resulting equation of motion,
\begin{equation}\label{Lphi3deformedeom}
\left(\square+m^2-g\phi\right)\left(\square\phi-\frac{g}{2}\phi^2\right)=0\,,
\end{equation}
takes the interesting form of an operator $(\square+m^2-g\phi )$ acting on the usual two-derivative equation of motion for massless $\phi^3$ theory. As we have seen before, the free equation of motion is $\square(\square+m^2)\phi^{(0)}=0$.

First we consider amplitudes involving only massless planewaves.  Any perturbative solution to the equation of motion of the  $\phi^3$ theory, $\square\phi=\frac{g}{2}\phi^2$, is clearly also a solution to the four-derivative equation of motion~(\ref{Lphi3deformedeom}).  Therefore, provided that the external-state boundary conditions $\phi^{(0)}$ are identified with planewaves, it must follow that the tree amplitudes in the four-derivative and two-derivative theories are trivially related,
\begin{align}\label{eq:Amp4der}
\cA^{(4)}_n(1,2,\ldots, n)=m^2 \cA^{(2)}_n(1,2,\ldots, n)\,.
\end{align}
The overall $m^2$ factor can be deduced from the $m\to\infty$ limit, where the four-derivative equation of motion simplifies to $m^2(\square\phi-\frac{g}{2}\phi^2)=0$.  For instance, at four points the planewave amplitude in the four-derivative theory is\footnote{We are omitting the overall momentum-conserving factor $(2\pi)^4\delta^{(4)}\left(\sum_ip_i\right)$, as we will do for all amplitudes in the remaining part of this paper.}
\begin{equation}
\cA^{(4)}_4(1,2,3,4)=
-im^2g^2\left(\frac{1}{s}+\frac{1}{t}+\frac{1}{u}\right)\,.
\end{equation}
where $s=(p_1\!+\!p_2)^2$, $t=(p_2\!+\!p_3)^2$ and $u=(p_1\!+\!p_3)^2$ --- this is indeed proportional to the usual $\phi^3$ amplitude.  In the $m\to0$ limit this amplitude vanishes, as do all planewave tree-level amplitudes of this (finely tuned) four-derivative theory. In the forthcoming sections we will see examples of gauge and gravity four-derivative theories that have exactly the same peculiar behavior. 

To obtain non-vanishing amplitudes in the massless limit we need to scatter one or more non-planewave states.  Let us focus on a single non-planewave state, and obtain the amplitude starting from the mass-deformed theory. It turns out that in the mass-deformed theory all amplitudes involving a single massive planewave vanish:
\begin{align}
\cA^{(4)}_n(1_m,2,\ldots, n)=0\,,
\end{align}
{where $1_m$ indicates that particle 1 is massive.} One can see this by considering the factorization channels of the tree amplitudes with only massless planewaves: as we have already discussed, these are proportional to those of the two-derivative $\phi^3$ theory. Hence there are no massive poles and therefore no residues corresponding to amplitudes with one massive external state.  

{Next, using the definition of non-planewave amplitudes~(\ref{eq:onenpw}), the above collapses to the following simple relation between four- and two-derivative amplitudes (particle $1$, the non-planewave mode, is denoted as $\tilde{1}$):}
\begin{align}\label{rel2}
\cA_n^{(4)}(\tilde{1},2,\ldots, n)=-\cA^{(2)}_n(1,2,\ldots, n)\,.
\end{align}
Note that both sides of this relation are supported on the usual momentum-conserving delta function (i.e. not a derivative of the delta function).  Since this relation has no overall factor of $m^2$ it implies that in the $m\to\infty$ limit the non-planewave amplitudes remain non-zero, and thus we have arrived at non-vanishing tree amplitudes in the massless four-derivative theory. 

We will not dwell further on amplitudes in the scalar model, as in this paper the focus is on gauge and gravity theories. However, we should emphasize that the relationships (\ref{eq:Amp4der}) and  (\ref{rel2}) between two- and four-derivative amplitudes apply more generally to other theories --- one simply requires that the equations of motion in a four-derivative theory be implied by those of a two-derivative theory. Conformal supergravity amplitudes with this behavior, at three points, have already been presented in refs.~\cite{Brodel:2009ep,Adamo:2018srx}.

Although this four-derivative toy model may seem rather contrived, all of the special features demonstrated here apply also to minimal conformal supergravity --- to be discussed in section~\ref{sec:mintheories}.  Of course, one can also add more interactions to the toy Lagrangian~(\ref{Lphi3deformed}) and obtain amplitudes unrelated to those of $\phi^3$ theory.  As we shall see in section~\ref{sec:confgravity},  gravitational analogies to such deformations of the interactions correspond to non-minimal versions of conformal supergravities. Before we get to conformal gravity, we will review a four-derivative gauge theory that features in the double copy of certain non-minimal conformal gravities.

%**************************************************
\section{Review of conformal gravity double copy}
\label{sec:review}

Scattering amplitudes in non-minimal $\cN=4$ Berkovits-Witten (BW) conformal supergravity (CSG) can be obtained from the following double copy~\cite{Johansson:2017srf}:
\begin{equation}
(\cN=4\text{ BW CSG}) = (DF)^2 \otimes (\cN=4\text{ SYM})\,,
\end{equation}
where $(DF)^2$ theory is shorthand for a certain four-derivative gauge theory built out of dimension-six operators, and the second gauge theory is $\cN=4$ super Yang-Mills (SYM). In this section we review the construction starting with the $(DF)^2$ Lagrangian, then give details of the double copy and finally consider natural deformations and extensions.

\subsection{The \texorpdfstring{$(DF)^2$}{DF2} theory}
\label{sec:nonminDF2}
The $(DF)^2$ theory is a bosonic $D$-dimensional gauge theory that is built entirely out of dimension-six operators (in $D=6$ counting); its Lagrangian is~\cite{Johansson:2017srf}
\begin{equation}
  \label{eq:DF2}
  \mathcal{L}_{(DF)^2}=\frac{1}{2}\left(D_{\mu}F^{\mu\nu,a}\right)^2-\frac{g}{3}F^3+\frac{1}{2}\left(D_{\mu}\varphi^\alpha\right)^2+\frac{g}{2}C^{\alpha ab}\varphi^\alpha F^{a}_{\mu\nu}F^{\mu\nu,b}+\frac{g}{6}d^{\alpha\beta\gamma}\varphi^\alpha\varphi^\beta\varphi^\gamma\,,
\end{equation}
where the field strength and covariant derivatives are defined as\footnote{Our Lie algebra generators and structure constants are defined such that $[T^a,T^b]=if^{abc}T^c$ and $\Tr(T^aT^b)=(1/2)\delta^{ab}$ for fundamental-representation generators.}
\begin{align}\label{eq:fieldstrengthdef}
\begin{aligned}
  F_{\mu\nu}^a &=\partial_\mu A_\nu^a-\partial_\nu A_\mu^a+gf^{abc}A_\mu^b A_\nu^c,\\
  D_\rho F_{\mu\nu}^a &=\partial_\rho F_{\mu\nu}^a+gf^{abc}A_\rho^b F_{\mu\nu}^c,\\
  F^3 &=f^{abc}{F_\mu}^{\nu,a}{F_\nu}^{\lambda,b}{F_\lambda}^{\mu,c},\\
  D_\mu\varphi^\alpha &=\partial_\mu\varphi^\alpha-ig(T^a_\mathsf{R})^{\alpha\beta}A_\mu^a\varphi^\beta,
  \end{aligned}
\end{align}
The field content comprises a gauge field $A_\mu^a$ transforming in the adjoint representation of a gauge group $G$, and a scalar field $\varphi^\alpha$ transforming in a real representation $\mathsf{R}$, with symmetric generators $(T^a_\mathsf{R})^{\alpha\beta}$.  It is assumed that the representation $\mathsf{R}$ appears in the tensor product of two adjoint representations, which defines the symmetric Clebsch-Gordan coefficients $C^{\alpha ab}=C^{\alpha ba}$.  In addition, the tensor product $\mathsf{R}\times \mathsf{R}$ contains $\mathsf{R}$ and this defines the totally symmetric $d^{\alpha\beta\gamma}$ tensor. The details of the representation are unimportant for calculating tree-level amplitudes; however, to ensure gauge invariance the Clebsch-Gordan coefficients need to transform as covariant tensors of a Lie algebra, which enforces the relations
\begin{align}
\begin{aligned}
  if^{abc}(T^c_\mathsf{R})^{\alpha\beta}&=(T^a_\mathsf{R})^{\alpha\gamma}(T^b_\mathsf{R})^{\gamma\beta}-(T^b_\mathsf{R})^{\alpha\gamma}(T^a_\mathsf{R})^{\gamma\alpha},\\
  iC^{\beta ab}(T^c_\mathsf{R})^{\alpha\beta} &=C^{\alpha ae}f^{bce}+C^{\alpha be}f^{ace},\\
  0 &=(T^a_\mathsf{R})^{\alpha\delta}d^{\delta\beta\gamma}+(T^a_\mathsf{R})^{\beta\delta}d^{\delta\gamma\alpha}+(T^a_\mathsf{R})^{\gamma\delta}d^{\delta\alpha\beta}\,.
  \end{aligned}
\end{align}

The kinetic term $(D_{\mu}F^{\mu\nu,a})^2$  has four derivatives, which implies that on the support of Lorenz gauge $\partial^\mu A^a_\mu=0$ the linearized equation of motion for the gauge field becomes $\square^2 A^a_\mu =0$.  This implies the existence of four independent spin-1 modes: two physical planewave gluons $A^\pm_\mu(p) = \eps^\pm_\mu(p)e^{ip\cdot x}$, and two ghost-like non-planewave gluons $\tilde A^\pm_\mu(p) =\eps^\pm_\mu(p) a_\nu x^\nu e^{ip\cdot x}$. Additionally, the field equation permits a ghost-like scalar excitation $A^0_\mu(p) = \eps^0_\mu(p)e^{ip\cdot x}$, which can be obtained as a longitudinal mode of a massive gluon after deforming the theory with a mass term (as done in \sec{sec:defsextensions}).  As explained for the toy model in section~\ref{sec:onshellstates}, one may calculate amplitudes involving physical gluons by isolating $1/p^4$ poles on external legs. We will in this section only consider physical gluons (see \sec{sec:minDF2} for non-planewave gluons).

Gluon tree amplitudes are easier to work with if we consider a color decomposition into color-ordered amplitudes for the gauge group $G=SU(N_c)$,
\begin{align}
  \cA^{(DF)^2}_n=g^{n-2}%(2\pi)^4\delta^{(4)}\Big(\sum_i p_i\Big)\!
  \sum_{\sigma\in S_n/Z_n}
\text{Tr}(T^{a_{\sigma_1}}T^{a_{\sigma_2}}\cdots T^{a_{\sigma_n}})
A_n^{(DF)^2}(\sigma_1,\sigma_2,\cdots,\sigma_n).
\end{align}
To obtain such a decomposition the color tensors $C^{\alpha ab}$ and $d^{\alpha\beta\gamma}$ can be massaged via
\begin{align}
\begin{aligned}
  C^{\alpha ab}C^{\alpha cd} &=f^{ace}f^{edb}+f^{ade}f^{ecb}\,,\\
  C^{\alpha ab}d^{\alpha\beta\gamma} &=(iT^a_\mathsf{R})^{\beta\alpha}(iT^b_\mathsf{R})^{\alpha\gamma}+(iT^b_\mathsf{R})^{\beta\alpha}(iT^a_\mathsf{R})^{\alpha\gamma}+C^{\beta ac}C^{\gamma cb}+C^{\beta bc}C^{\gamma ca}\,,
  \end{aligned}
  \label{Cdconsistency}
\end{align}
until every color factor in the tree amplitude can be expressed in terms of only $f^{abc}$ structure constants, at which point they can be converted to the trace basis through standard SU($N_c$) identities. 
The two relations (\ref{Cdconsistency}) are also necessary~\cite{Johansson:2017srf} for the color-ordered amplitudes with physical gluons to satisfy BCJ tree-amplitude relations~\cite{Bern:2008qj}. Since the $(DF)^2$ theory satisfies color-kinematics duality, at least through eight points~\cite{Johansson:2017srf}, we may employ it to construct gravitational amplitudes via the double copy.

\subsection{Double copy with \texorpdfstring{$\cN=4$}{N=4} SYM}

First we discuss the double copy at the level of the physical on-shell states that we are interested in scattering. The complete $\cN=4$ SYM on-shell supermultiplet is
\begin{align}\label{eq:neq4multiplet}
\cV=A^++\eta^I\lambda^+_I+\frac{1}{2}\eta^I\eta^JS_{IJ}+\frac{1}{3!}\eps_{IJKL}\eta^I\eta^J\eta^K\lambda_-^L+\eta^1\eta^2\eta^3\eta^4A^-.
\end{align}
where $I,J,\ldots$ are SU(4) R-symmetry indices and $\eta^I$ are Grassmann-odd auxiliary parameters used in the on-shell superspace formalism (see e.g.~\cite{Elvang:2015rqa}).  The on-shell particle content resulting from a double copy of physical $(DF)^2$ gluon states, $A^+$ and $A^-$, with those of $\cN=4$ SYM coincides with that of chiral and anti-chiral graviton supermultiplets in $\cN=4$ conformal supergravity:
\begin{align}\label{eq:neq4sgmultiplet}
  \begin{aligned}
   \cH^+&\equiv\! A^+ \otimes \cV=h^{++}+\eta^I\psi^+_I+\frac{1}{2}\eta^I\eta^JA^+_{IJ}+\frac{1}{3!}\eps_{IJKL}\eta^I\eta^J\eta^K\Lambda_+^L+\eta^1\eta^2\eta^3\eta^4\,\bC,\\
   \cH^- &\equiv\! A^- \otimes \cV =C+\eta^I\Lambda^-_I+\frac{1}{2}\eta^I\eta^JA^-_{IJ}+\frac{1}{3!}\eps_{IJKL}\eta^I\eta^J\eta^K\psi_-^L+\eta^1\eta^2\eta^3\eta^4\,h^{--}.
  \end{aligned}
\end{align}
These are the same on-shell graviton supermultiplets as are relevant for $\cN=4$ Einstein supergravity.  The additional ghost-like states that are prolific in $\cN=4$ conformal supergravity can in principle be considered (since they appear in factorization channels of the tree-level amplitudes); however, we will not do so here. 

For tree-level amplitudes involving only external graviton multiplets, it is convenient to phrase the BCJ double copy~\cite{Bern:2008qj,Bern:2010ue} in terms of the KLT formula~\cite{Kawai:1985xq}. Doing this for the graviton multiplets in \eqn{eq:neq4sgmultiplet} gives the following formula for Berkovits-Witten conformal supergravity tree amplitudes:
\begin{align}\label{eq:doublecopyCG}
M_n^{\text{BW CSG}}=
\sum_{\sigma,\rho \in S_{n-3}}A_n^{(DF)^2}(1,\sigma,n,n-1)\mathcal{S}[\sigma|\rho]A_n^\text{SYM}(1,\rho,n-1,n)\,, 
\end{align}
where we have suppressed the gravitational coupling constant. The KLT kernel $\mathcal{S}[\sigma|\rho]$ is an $(n-3)! \times (n-3)!$ matrix of kinematic polynomials that acts on the color-ordered amplitudes for $(n-3)!$ permutations of the external legs~\cite{Bern:1998sv,BjerrumBohr:2010ta,BjerrumBohr:2010zb,BjerrumBohr:2010yc,BjerrumBohr:2010hn}:
\begin{align}
  \mathcal{S}[\sigma|\rho]=\prod_{i=2}^{n-2}\left[2p_1\cdot p_{\sigma_i}+\sum_{j=2}^{i}2p_{\sigma_i}\cdot p_{\sigma_j}\theta(\sigma_j,\sigma_i)_\rho\right]\,,
\end{align}
where $\theta(\sigma_j,\sigma_i)_\rho=1$ if $\sigma_j$ is before $\sigma_i$ in the permutation $\rho$, and zero otherwise.

As is manifest in the double copy, the four-dimensional conformal supergravity amplitudes are classified according to (i) how many external states belong to the $\cH^+$ or $\cH^-$ multiplets  (inherited from the helicity of the $(DF)^2$ side), and (ii) their maximally-helicity-violating (N$^k$MHV) degree (inherited from the $\cN=4$ SYM side).  The MHV-sector amplitudes were first computed by Berkovits and Witten in ref.~\cite{Berkovits:2004jj}, and have since been confirmed up to eight points using the above double copy~\cite{Johansson:2017srf}. They are all given by the compact formula
\begin{align}\label{eq:BW}
M^\text{BW CSG}_n(\cH_1^+,\cdots,\cH_{k}^+,\cH_{k+1}^-,\cdots,\cH_n^-)=
(-1)^ni\delta^8(Q)\prod_{i=1}^k\sum_{j=1,j\neq i}^n
\frac{[ij]\braket{jq}^2}{\braket{ij}\braket{iq}^2},
\end{align}
where $q$ is a reference choice and $\delta^{8}(Q)=\delta^{8}(\sum_i \lambda_i^{\alpha} \eta^I_i)$ is the usual supermomentum-conserving delta function given in terms of on-shell spinors $\lambda_i^{\alpha}$ and the Grassmann-odd numbers~\cite{Nair:1988bq}.  Historically, these amplitudes arose from the $\cN=4$ twistor string theory as a conformal gravity ``contamination'' of $\cN=4$ SYM in the multi-trace sector~\cite{Berkovits:2004jj} (see also refs.~\cite{Dolan:2008gc,Adamo:2013tja,Farrow:2018yqf}). Although some hints were provided in Berkovits and Witten's original work, a Lagrangian formulation of the field theory from which these amplitudes originate has until now been lacking.  This will be discussed in section~\ref{sec:confgravity}, where a Lagrangian is presented.

\subsection{Deformations and extensions}
\label{sec:defsextensions}

Several extensions to the $(DF)^2$ theory were considered in ref.~\cite{Johansson:2017srf}.  Firstly, the theory was mass deformed by adding a Yang-Mills term to the Lagrangian~\eqref{eq:DF2}, thus resolving the states coming from $A_\mu^a$ into massless and massive gluon states (five states in total).  For the resulting theory to still obey color-kinematics duality, the scalar $\varphi^\alpha$ should also acquire the same mass.  The mass-deformed Lagrangian is~\cite{Johansson:2017srf}
\begin{align}\label{eq:df2MassDef}
  \mathcal{L}_{(DF)^2+\text{YM}}=\mathcal{L}_{(DF)^2}-\frac{m^2}{4}(F^a_{\mu\nu})^2-\frac{m^2}{2}\varphi^\alpha\varphi^\alpha,
\end{align}
and $m$ effectively interpolates between the $(DF)^2$ and YM theories in the $m\to0$ and $m\to\infty$ limits respectively.  Upon taking the double copy~\eqref{eq:doublecopyCG} between $(DF)^2+ \rm{YM}$ and SYM, the resulting gravitational amplitudes interpolate between Berkovits-Witten amplitudes and those of Einstein supergravity.

Another deformation considered in ref.~\cite{Johansson:2017srf} was the inclusion of a bi-adjoint scalar $\phi^{aA}$ with cubic self interactions.  This scalar transforms in the adjoint of the gauge group $G$, and also in the adjoint of a global group $\tilde G$.  The complete Lagrangian in this case is
\begin{align}
  \label{eq:DF2YMphi3}
  \mathcal{L}_{(DF)^2+\text{YM}+\phi^3}&=\mathcal{L}_{(DF)^2}-\frac{m^2}{4}(F^a_{\mu\nu})^2-\frac{m^2}{2}\varphi^\alpha\varphi^\alpha+\frac{1}{2}\left(D_\mu\phi^{aA}\right)^2\nonumber\\
   &\quad+\frac{g}{2}C^{\alpha ab}\varphi^\alpha\phi^{aA}\phi^{bA}+\frac{g\lambda}{3!}f^{abc}\hat{f}^{ABC}\phi^{aA}\phi^{bB}\phi^{cC}\,,
\end{align}
where $\hat{f}^{ABC}$ are the structure constants of $\tilde G$ and the coupling $\lambda$ is a free parameter.  The coupling to $\varphi^\alpha$ through $C^{\alpha ab}$ is necessary to ensure valid BCJ amplitude relations when the bi-adjoint scalars $\phi^{aA}$ are scattered.  This theory plays a crucial role in the novel double-copy constructions of heterotic and bosonic string amplitudes~\cite{Azevedo:2018dgo}. In the $m\to0$ limit, one obtains a $(DF)^2+\phi^3$ theory, and the resulting  amplitudes from the double copy with $\cN=4$ SYM give rise to additional vector multiplets in conformal gravity.

Finally, in the $m\to\infty$ limit of this deformation the kinetic term of $\varphi^\alpha$ drops out.  This scalar may therefore be integrated out, giving rise to a $\phi^4$ interaction.  Keeping only terms proportional to $m^2$, the resulting Lagrangian becomes~\cite{Chiodaroli:2014xia}
\begin{align}\label{eq:YMPhi3Lagrangian}
\begin{aligned}
\mathcal{L}_{\text{YM}+\phi^3}&=
-\frac{1}{4}(F^a_{\mu\nu})^2+\frac{1}{2}\left(D_\mu\phi^{aA}\right)^2\\
&\qquad+\frac{g\lambda'}{3!}f^{abc}\hat{f}^{ABC}\phi^{aA}\phi^{bB}\phi^{cC}-
\frac{g^2}{4}f^{ace}f^{ebd}\phi^{aA}\phi^{bA}\phi^{cB}\phi^{dB},
\end{aligned}
\end{align}
where the overall $m^2$ has been removed by rescaling, and $\lambda'=m\lambda$ is kept finite in the limit.  To realize this explicitly, one should make the field $\phi^{aA}$ dimension-one by rescaling $\phi^{aA}\to m\phi^{aA}$.  Taking the double copy with $\cN=4$ SYM amplitudes now gives $\cN=4$ Einstein-Yang-Mills (EYM) supergravity amplitudes.  More generally, by varying the degree of supersymmetry one obtains~\cite{Chiodaroli:2014xia}
\begin{equation}
(\cN=0,1,2,4\text{ EYM SG}) = (\text{YM}+\phi^3) \otimes (\cN=0,1,2,4\text{ SYM})\,.
\end{equation}
Both Einstein and EYM supergravity theories are discussed in the next section.

%**************************************************
\section{Einstein supergravity}
\label{sec:einsteingravity}

To build familiarity with the formalism used to describe gravitational theories in this paper we first discuss some details of four-dimensional $\cN=4$ Einstein supergravity.  For simplicity of presentation, we will focus on the fields that are singlets under the SU(4) R-symmetry of the theory, as this is often sufficient to reconstruct the tree amplitudes for all other fields (e.g. through supersymmetric Ward identities~\cite{Elvang:2009wd}). 

The SU(4)-singlet sector of the pure $\cN=4$ theory --- which is sometimes playfully referred to as $\cN=0$ supergravity --- consists of ordinary Einstein gravity coupled to a scalar dilaton $\varphi$ and a pseudoscalar axion $\chi$.\footnote{The four-dimensional axion $\chi$ emerges from the $D$-dimensional antisymmetric $B_{\mu\nu}$ tensor as $H_{\mu\nu\rho}=\frac{i}{2}e^{\kappa\varphi}e\eps_{\mu\nu\rho\sigma}\partial^\sigma\chi$, where $H_{\mu\nu\rho}=\nabla_\mu B_{\nu\rho}+\text{cyclic}$ is the curvature of $B_{\mu\nu}$ (see e.g. ref.~\cite{Schwarz:1992tn}).}  Its Lagrangian is
\begin{align}\label{eq:nmGravity}
e^{-1}\cL=-\frac{2}{\kappa^2}R+\frac{1}{\kappa^2}
\left(\partial_\mu\varphi\partial^\mu\varphi+e^{2\varphi}\partial_\mu\chi\partial^\mu\chi\right),
\end{align}
where $e=\sqrt{-g}$, and $g$ is the determinant of the metric (not to be confused with the gauge coupling constant). The gravitational coupling $\kappa$ is made explicit here, but in the remaining part of this section we set $\kappa=2$.\footnote{With respect to the complete $\cN=4$ Lagrangian given in ref.~\cite{Cremmer:1977tt} we set $K=\frac{1}{2}$.}

The SU(4)-singlet part of the $\cN=4$ Lagrangian may be more compactly expressed using the complex scalar field $\tau$,
\begin{align}
\tau=\chi+ie^{-\varphi},
\end{align}
giving
\begin{align}\label{eq:nmGravityTau}
e^{-1}\cL=-\frac{R}{2}+\frac{\partial_\mu\bar{\tau}\partial^\mu\tau}{4(\text{Im}\,\tau)^2}.
\end{align}
Although the origin of this four-dimensional field is different, the way $\tau$ appears in the Lagrangian should be familiar from the ten-dimensional effective actions of string theory~\cite{Polchinski:1998rr}.

\subsection{Covariant SU(1,1)/U(1) formulation}\label{sec:cosetformulation}

It is well-known that the scalar sector of $\cN=4$ supergravity~(\ref{eq:nmGravityTau}) realizes a global nonlinear SL(2,$\mathbb{R}$)$\cong$SU(1,1) symmetry:
\begin{align}
\tau\to\frac{a\tau+b}{c\tau+d}, &&
\text{det}\left(\begin{matrix} a && b \\ c && d \end{matrix}\right)=1, &&
a,b,c,d\in\mathbb{R}.
\end{align}
Any value of $\tau$ is also invariant under a U(1) stabilizer subgroup (which also acts on the additional matter content in the theory), so one typically regards $\tau$ as living in an SU(1,1)/U(1) coset space.  

As the SU(1,1) symmetry is an important feature of the theories considered in this paper, it would be advantageous to make it act linearly on the fields. Such a linear realization is achieved by moving to the covariant formulation.  We introduce a doublet of complex scalars $\phi^\alpha$ ($\alpha=1,2$), subject to constraints
\begin{align}
\phi^\alpha\phi_\alpha=1, && \phi_1=\bar{\phi}^1, && \phi_2=-\bar{\phi}^2.
\end{align}
The linear action of $U\in$ SU(1,1) on these scalars is simply $\phi'^\alpha={U^\alpha}_\beta\phi^\beta$.  Meanwhile, the U(1) stabilizer is promoted to a local symmetry, acting on the scalars with weight 1, i.e.~$\phi^\alpha\to e^{i\lambda_A(x)}\phi^\alpha$.  To write down U(1)-covariant derivatives $\tilde{\nabla}_\mu$ we use a composite gauge field $a_\mu=i\phi^\alpha\partial_\mu\phi_\alpha$ --- so, for instance, $\tilde{\nabla}_\mu\phi^\alpha=(\partial_\mu-ia_\mu)\phi^\alpha$ also transforms with weight 1.  Finally, we define the SU(1,1)/U(1) coset fields
\begin{align}
P_\mu=\phi^\alpha\eps_{\alpha\beta}\tilde{\nabla}_\mu\phi^\beta, &&
\bP_\mu=-\phi_\alpha\eps^{\alpha\beta}\tilde{\nabla}_\mu\phi_\beta,
\end{align}
with $\eps_{12}=\eps^{12}=1$.  These are invariant under SU(1,1) and carry U(1) weights $+2$ and $-2$ respectively.  In terms of them, the SU(4)-singlet part of ${\cal N}=4$ Einstein supergravity Lagrangian is simply
\begin{align}\label{eq:nmGravityP}
e^{-1}\cL=-\frac{R}{2}+P\cdot\bP.
\end{align}

Of course, these constrained scalars are unsuitable for scattering.  So we parametrize them in terms of unconstrained scalars, and the covariant formulation gives us flexibility in how we do this.  For instance, the U(1) gauge choice $\text{Im}(\phi^1+\phi^2)=0$ can be parametrized as~\cite{Ferrara:2012ui}
\begin{align}\label{eq:tauParam}
\phi^1=\frac{1}{2\sqrt{\text{Im}\,\tau}}(1-i\tau), &&
\phi^2=\frac{1}{2\sqrt{\text{Im}\,\tau}}(1+i\tau), &&
\tau=i\frac{\phi^1-\phi^2}{\phi^1+\phi^2},
\end{align}
which solves $\phi^\alpha\phi_\alpha=1$.  Via the definitions of $P_\mu$ and $a_\mu$ given above this implies
\begin{align}\label{eq:tauCosets}
P_\mu=\frac{i\partial_\mu\tau}{2\text{Im}\,\tau}=\frac{1}{2}(\partial_\mu\varphi+ie^\varphi\partial_\mu\chi), &&
a_\mu=-\frac{\partial_\mu(\tau+\bar{\tau})}{4\text{Im}\,\tau}=-\frac{1}{2}e^\varphi\partial_\mu\chi,
\end{align}
thereby confirming that the two versions of the ${\cal N}=4$ supergravity Lagrangian~(\ref{eq:nmGravityTau}) and~(\ref{eq:nmGravityP}) are equivalent.

Since $\tau$ should be perturbatively expanded around the point $\braket{\tau}=i$, it is somewhat inconvenient to consider amplitudes in terms of this field.  Following ref.~\cite{Bergshoeff:1980is}, an alternative useful gauge fixing is the reality condition $\phi^1=\phi_1$, parametrized as
\begin{align}\label{eq:cayleighParam}
\phi^1=\frac{1}{\sqrt{1-|C|^2}}, && \phi^2=-\frac{C}{\sqrt{1-|C|^2}}, && C=-\frac{\phi^2}{\phi^1}.
\end{align}
In this case, one can easily show that
\begin{align}\label{eq:cayleighCosets}
P_\mu=-\frac{\partial_\mu C}{1-|C|^2}, && a_\mu=\frac{\text{Im}(C\partial_\mu\bar{C})}{1-|C|^2}.
\end{align}
Notice that these expressions for $P_\mu$ and $a_\mu$ do not equal the ones given above in terms of $\tau$~(\ref{eq:tauCosets}); the former are obtained from the latter via the U(1) transformation~\cite{Chiodaroli:2009yw}
\begin{align}
P_\mu\to e^{2i\theta}P_\mu, && a_\mu\to a_\mu +\partial_\mu\theta, &&
\theta=\frac{1}{2i}\text{log}\left(\frac{1-i\tau}{1+i\bar{\tau}}\right).
\end{align}
$C$ transforms non-linearly under SU(1,1), and the local U(1) symmetry is broken to a global, chiral U(1) under which $C$ has a charge of $2$.

\subsection{Double copy structure}

With the parametrization choice (\ref{eq:cayleighCosets}),  the SU(4)-singlet part of the ${\cal N}=4$ supergravity Lagrangian~(\ref{eq:nmGravityP}) becomes
\begin{align}
e^{-1}\cL=-\frac{R}{2}+\frac{\partial_\mu\bar{C}\partial^\mu C}{(1-|C|^2)^2}.
\end{align}
Using this Lagrangian as a starting point, amplitude calculations involving gravitons and scalars are, in principle, straightforward.  One simply expands $g_{\mu\nu}=\eta_{\mu\nu}+\kappa h_{\mu\nu}$ and reads off interaction vertices and propagators in the usual way (see e.g. ref.~\cite{Weinzierl:2016bus}).  The denominator involving $C$ expands to give an infinite tower of interactions with itself, its conjugate $\bC$, and the graviton $h_{\mu\nu}$ field.

As is well known, and as we have confirmed by direct calculations, the resulting graviton-scalar tree amplitudes coincide with those computed from the double copy~\cite{Bern:2011rj,Carrasco:2013ypa}:
\begin{align}
(\cN=0,1,2,4\text{ Einstein SG}) = {\rm YM} \otimes (\cN=0,1,2,4\text{ SYM})\,.
\end{align}
Here the $\cN<4$ supergravity theories are non-pure, in the sense that they have additional complex matter multiplets where the scalars $C$, $\bC$ are the top and bottom components, respectively.  As should be familiar from the  $\cN=0$ case of the double copy, we can identify the linearized on-shell dilaton $\varphi$ and axion $\chi$ states with symmetric and antisymmetric combinations of on-shell gluon states, respectively,
\begin{align}
\varphi=-A^+\otimes A^--A^-\otimes A^+, && i\chi=A^+\otimes A^--A^-\otimes A^+.
\end{align}

Using the non-linear expressions for $\tau$ and $C$ in terms of $\phi^\alpha$ (eqs.~(\ref{eq:tauParam}) and (\ref{eq:cayleighParam}) respectively), one can express $C$ in terms of $\varphi$ and $\chi$:
\begin{align}
C=-\frac{1+i\tau}{1-i\tau}=
-\frac{1-e^{-\varphi}+i\chi}{1+e^{-\varphi}-i\chi}=
-\frac{1}{2}(\varphi+i\chi)+\cdots.
\end{align}
Therefore, at the linearized on-shell level one can also identify
\begin{align}
C=A^-\otimes A^+, && \bC=A^+\otimes A^-,
\end{align}
which justifies our use of the $C$ and $\bC$ states in the $\cH^-$ and $\cH^+$ multiplets~(\ref{eq:neq4sgmultiplet}) respectively.

We mentioned earlier that $C$ has a U(1) charge of $2$.  The U(1) charges of other states in the $\cH^+$ and $\cH^-$ multiplets are given by the difference of their helicities on the two sides of the double copy: $q_{\text{U(1)}}=h(\text{SYM})-h(\text{YM})$.  When computing tree-level amplitudes this symmetry is useful --- for instance, it explains the decoupling of the scalars in pure-graviton tree amplitudes. Since the work of Marcus~\cite{Marcus:1985yy}, it is known that the U(1) symmetry of $\cN=4$ Einstein supergravity is anomalous at the quantum level (see refs.~\cite{Carrasco:2013ypa,Bern:2017tuc,Bern:2017rjw} for studies of the anomaly from the double-copy perspective).  U(1) symmetry breaking is also an important feature of non-minimal conformal supergravities, where it happens already at tree level~\cite{Fradkin:1985am}.

\subsection{Coupling to vector multiplets}

Before proceeding to conformal gravity we review the coupling of additional $\cN=4$ vector multiplets to Einstein supergravity.  With a vector multiplet that transforms in the adjoint of a gauge group $\tilde G$, the Lagrangian~(\ref{eq:nmGravityP}) generalizes to that of $\cN=4$ Einstein-Yang-Mills (EYM) supergravity. The SU(4)-singlet part is
\begin{align}\label{eq:neq4PSGwYM}
e^{-1}\mathcal{L}&=-\frac{R}{2}+
P\cdot\bP-\frac{1}{4}\big[i\bar{\tau}(F^{+,A}_{\mu\nu})^2-i\tau(F^{-,A}_{\mu\nu})^2\big],
\end{align}
where we have introduced the (anti-)self-dual part of the Yang-Mills field strength
\begin{align}\label{eq:selfdualCurvature}
F^{\pm,A}_{\mu\nu}=\frac{1}{2}\left(F^A_{\mu\nu}\pm\tilde{F}^A_{\mu\nu}\right), &&
\tilde{F}^A_{\mu\nu}=\frac{i}{2}e\epsilon_{\mu\nu\rho\sigma}F^{\rho\sigma,A}.
\end{align}
Here $F_{\mu\nu}^A =\partial_\mu A_\nu^A-\partial_\nu A_\mu^A+\lambda \hat f^{ABC}A_\mu^B A_\nu^C$, and $\lambda$ is the coupling constant that also appears in the deformed $(DF)^2$ theory (\ref{eq:DF2YMphi3}).
Writing the EYM Lagrangian explicitly in terms of the dilaton $\varphi$ and axion $\chi$, it becomes
\begin{align}\label{eq:neq4PSGwVectors}
e^{-1}\mathcal{L}=-\frac{R}{2}+\frac{1}{4}\left(\partial_\mu\varphi\partial^\mu\varphi\!+\!
e^{2\varphi}\partial_\mu\chi\partial^\mu\chi\!-\!e^{-\varphi}F^A_{\mu\nu}F^{\mu\nu,A}\!-\!
i\chi F^A_{\mu\nu}\tilde{F}^{\mu\nu,A}\right).
\end{align}

The MHV single-trace amplitudes --- coefficient of $\lambda^{n-k-1}\text{Tr}(T^{A_k}\cdots T^{A_n})$ --- in this theory admit a simple form.\footnote{When writing color-stripped amplitudes we ignore an overall factor of $\sqrt{2}$ for each gluon leg.}  If two or more external states belong to the $\cH^-$ multiplet~(\ref{eq:neq4sgmultiplet}) then the amplitudes vanish.  Otherwise~\cite{Bern:1999bx,Feng:2012sy,Du:2016wkt}
\begin{align}\label{eq:mhvEYMSG}
M_n(\cH_1^+\cdots\cH_{k-2}^+\cH_{k-1}^\pm\cV_k\cdots\cV_n)=
%g^{n-k-1}\left(\frac{\kappa}{2}\right)^{k-1}\!\!\!\!
i \left(-1\right)^{k-1}\frac{\delta^8(Q)}{\braket{k,k\!+\!1}\cdots\braket{nk}}\det\left(\Phi^+\right),
\end{align}
where $\Phi^+$ is the $(k\!-\!2)$- or $(k\!-\!1)$-dimensional Hodges' matrix~\cite{Hodges:2011wm,Hodges:2012ym}, implemented on the $\cH^+$ states as
\begin{align}
(\Phi^+)_i^j=-\frac{[ij]}{\braket{ij}} \qquad \text{for } i\neq j, &&
(\Phi^+)_i^i=\sum_{j=1,j\neq i}^n\frac{[ij]\braket{jx}\braket{jy}}{\braket{ij}\braket{ix}\braket{iy}}.
\end{align}
By taking only the diagonal elements of Hodges' matrix and setting $q=x=y$ one precisely reproduces the factor in the Berkovits-Witten amplitude~(\ref{eq:BW}) associated with the $\cH^+$ states.  The MHV-sector double-trace amplitudes --- coefficient of $\lambda^{n-4}\text{Tr}(T^{A_1}\cdots T^{A_{r\!-\!1}})\text{Tr}(T^{A_r}\cdots T^{A_n})$ --- are given by~\cite{Cachazo:2014nsa}
\begin{align}\label{eq:mhvEYMSGDoubleTrace}
M_n(\cV_1\cdots \cV_{r-1}|\cV_r\cdots\cV_n)=
%g^{n-4}\left(\frac{\kappa}{2}\right)^2
-\frac{ip^2_{r,n}\delta^8(Q)}{\braket{12}\cdots\braket{r\!-\!1,1}\braket{r,r\!+\!1}\cdots\braket{nr}},
\end{align}
and $p_{i,j}=p_i+p_{i+1}+\cdots+p_j$.

As mentioned in section~\ref{sec:defsextensions}, these amplitudes admit a double copy construction from the ${\rm YM}+\phi^3$ theory defined by the Lagrangian~(\ref{eq:YMPhi3Lagrangian}).  We have checked the MHV-sector single-trace~(\ref{eq:mhvEYMSG}) and double-trace~(\ref{eq:mhvEYMSGDoubleTrace}) amplitudes by explicit calculation from the EYM Lagrangian~(\ref{eq:neq4PSGwYM}) for a variety of external states including scalars, gravitons and gluons.  We have also checked the single-, double-, and triple-trace sectors of the six-point NMHV amplitude $M_6(A^+,A^+,A^+,A^-,A^-,A^-)$ by comparison with the double copy.

%**************************************************
\section{Minimal conformal supergravity}
\label{sec:mintheories}

In this section we explore scattering amplitudes in minimal conformal supergravity.  We show that the amplitudes admit a double-copy construction from a truncation of the $(DF)^2$ theory reviewed in section~\ref{sec:nonminDF2}, which we therefore refer to as minimal $(DF)^2$ theory.  In general, we will consider mass-deformed versions of these theories, and thus the proper theories are obtained in the $m\rightarrow 0$ limit. 
By analogy to the four-derivative toy model discussed in section~\ref{sec:ttoymodel}, the four-derivative equations of motion in these theories are automatically solved by classical solutions to Yang-Mills and Einstein supergravity, respectively.  Hence their tree amplitudes for planewave states vanish for $m= 0$, and are equal (up to a factor of $m^2$) to their two-derivative counterparts for finite $m$.  The vanishing of planewave amplitudes in minimal conformal supergravity follows from an argument by Maldacena~\cite{Maldacena:2011mk}, which was later elaborated on in refs.~\cite{Adamo:2013tja,Adamo:2016ple,Beccaria:2016syk}.

\subsection{Minimal \texorpdfstring{$(DF)^2$}{DF2} theory}
\label{sec:minDF2}

We define the minimal $(DF)^2$ theory to consist solely of the kinetic term $\left(D_{\mu}F^{\mu\nu,a}\right)^2$ from the Lagrangian (\ref{eq:DF2}). However, it is convenient to introduce the Yang-Mills term from the very beginning, thus the mass-deformed minimal theory has the Lagrangian\footnote{Up to an overall $m^2$ factor, this is the Lee-Wick theory~\cite{Lee:1969fy,Lee:1970iw} considered in ref.~\cite{Grinstein:2007mp}.}
\begin{align}
\cL=\frac{1}{2}\left(D_{\mu}F^{\mu\nu,a}\right)^2-\frac{m^2}{4}(F^a_{\mu\nu})^2\,.
\end{align}
We note that the classical equations of motion,
\begin{align}
  D_\lambda D^\lambda D_\rho F^{\rho\mu,a}-D_\lambda D^\mu D_\rho F^{\rho\lambda,a}+[D^\mu,D^\lambda]D^\rho F^a_{\rho\lambda}+m^2 D_\rho F^{\rho\mu,a}=0\,,
\end{align}
are solved by the Yang-Mills equations $D_{\mu}F^{\mu\nu,a}=0$.  The situation is therefore completely analogous to the scalar toy model in \sec{sec:ttoymodel}. Tree amplitudes involving physical gluons $A^+$ and $A^-$ in the mass-deformed theory are proportional to those of ordinary Yang-Mills theory, 
\begin{align}\label{eq:pureDF2Amp}
A_n^{(4)}(1,2,\ldots,n)=m^2 A_n^{(2)}(1,2,\ldots,n)\,,
\end{align}
implying that they vanish identically in the minimal $(DF)^2$ theory ($m\rightarrow 0$). We remind the reader that the superscripts are used to distinguish the four- and two-derivative theories.  Also, tree amplitudes involving a single non-planewave state are given by
\begin{align}\label{eq:pureDF2Amp2}
A_n^{(4)}(\tilde{1},2,\ldots,n)=-A_n^{(2)}(1,2,\ldots,n)\,,
%\cA_n^{(4)}(\tilde{1},2,\ldots,n)=-(2\pi)^4\delta^{(4)}\Big(\sum_i p_i\Big)A_n^{(2)}(1,2,\ldots,n)\,.
\end{align}
and thus there exist non-vanishing amplitudes in the $m\rightarrow 0$ limit.

A minor difference compared to the scalar toy model is gauge fixing. It is convenient to use Lorenz gauge $\partial^\mu A^a_\mu=0$ in both the two- and four-derivative theories. In the latter case, the gauge-fixing term is $\cL_\text{GF}=-\frac{1}{2}(\partial_\mu A^{\mu,a})(\square+m^2)(\partial_\nu A^{\nu,a})$, which gives a simple propagator,
\begin{align}
\begin{tikzpicture}
  [baseline={([yshift=-.5ex]current bounding box.center)},thick,inner sep=0pt,minimum size=0pt,>=stealth,scale=0.5]
  \node (1) at (0,0) {};
  \node (2) at (4,0) {};
  \draw[gluon] (1) node[left=0.1] {$A^{\mu,a}$} to node[below=0.25] {$p$} (2) node[right=0.1] {$A^{\nu,b}$} {};
\end{tikzpicture}
=\frac{i\,\eta^{\mu\nu}\delta^{ab}}{p^2(p^2-m^2)}\,.
\end{align}
This works equally well when $m=0$, in which case the propagator is $i\eta^{\mu\nu}\delta^{ab}/p^4$.

\subsection{Weyl gravity}
\label{sec:weyl2}

The simplest four-dimensional conformal gravity is the pure Weyl theory. Including an Einstein-Hilbert term from the start gives the mass-deformed Weyl theory,
\begin{align}\label{eq:weylLagrangian}
e^{-1}\cL&=-\frac{1}{\varkappa^2}(W_{\mu\nu\rho\sigma})^2-\frac{2}{\varkappa^2}m^2R,
\end{align}
where $\varkappa$ is a dimensionless coupling.\footnote{Note that the ordinary gravitational coupling is given by $\kappa=\varkappa/m$, and for $\varkappa = {\cal O}(1)$ the Planck mass can be identified with $m$.} For simplicity, from now on we set $\varkappa=2$.  The four-dimensional Weyl tensor, and its square, are expressed as
\begin{align}\label{eq:weylTensor}
\begin{aligned}
W_{\mu\nu\rho\sigma}&=R_{\mu\nu\rho\sigma}+g_{\nu[\rho}R_{\sigma]\mu}-g_{\mu[\rho}R_{\sigma]\nu}+\frac{1}{3}g_{\mu[\rho}g_{\sigma]\nu}R,\\
(W_{\mu\nu\rho\sigma})^2&=(R_{\mu\nu\rho\sigma})^2-2(R_{\mu\nu})^2+\frac{1}{3}R^2=
2(R_{\mu\nu})^2-\frac{2}{3}R^2+\text{GB},
\end{aligned}
\end{align}
where GB is the topological Gauss-Bonnet term.  Using that the Weyl tensor with one raised index ${W^\mu}_{\nu\rho\sigma}$ is invariant under dilatations $g_{\mu\nu}\to e^{-2\lambda_D(x)}g_{\mu\nu}$, it follows that the massless theory has local scale symmetry.

A graviton field is obtained by considering the metric perturbation $g_{\mu\nu}=\eta_{\mu\nu}+\varkappa h_{\mu\nu}$ (instead of $\kappa$ as we used earlier).  The equations of motion are $B_{\mu\nu}+m^2R_{\mu\nu}=0$, where the Bach tensor $B_{\mu\nu}$ is
\begin{align}
\begin{aligned}
B_{\mu\nu}&=-(2\nabla^\rho\nabla^\sigma+R^{\rho\sigma})W_{\rho\mu\nu\sigma}\\
&=-2\nabla^\rho\nabla_{(\mu}{R_{\nu)\rho}}+\square R_{\mu\nu}+\frac{2}{3}\nabla_\mu\nabla_\nu R-
\frac{1}{6}g_{\mu\nu}\square R\\
&\quad+2R_{\rho\mu}{R^\rho}_\nu-\frac{2}{3}R_{\mu\nu}R-\frac{1}{2}g_{\mu\nu}\Big((R_{\rho\lambda})^2-\frac{1}{3}R^2\Big)\,.
\end{aligned}
\end{align}
Any solution to the vacuum Einstein equations $R_{\mu\nu}=0$ is also a solution to these equations of motion.  Therefore, as before, the following tree amplitudes of the four-derivative mass-deformed Weyl theory are related to those of Einstein gravity:
\begin{equation}
M^{(4)}_n(1,2,\ldots,n)=m^2M^{(2)}_n(1,2,\ldots,n)\,,~~~ M^{(4)}_n(\tilde{1},2,\ldots,n)=-M^{(2)}_n(1,2,\ldots,n)\,.
\end{equation}
This is strikingly analogous to Maldacena's argument on the relation between conformal gravity and Einstein gravity amplitudes~\cite{Maldacena:2011mk}; however, he considered curved AdS space where the scale was provided by $\Lambda_{\rm AdS }$ instead of $m^2$. 
The vanishing of planewave amplitudes in flat-space conformal gravity follows either way in the limits $\Lambda_{\rm AdS }, m^2 \rightarrow 0$~\cite{Adamo:2013tja,Adamo:2016ple,Beccaria:2016syk}.\footnote{An alternative argument, following 't Hooft and Veltman's classic work~\cite{tHooft:1974toh} (see also~\cite{Deser:1986xr}), is that $(R_{\mu\nu})^2$ and $R^2$ terms can be removed by a field redefinition $g_{\mu\nu} \rightarrow g_{\mu\nu} + a R_{\mu\nu}+ b g_{\mu\nu} R$, thus $(W_{\mu\nu\rho\sigma})^2 \sim {\rm GB}$  vanishes in four dimensions. Note, however, this argument assumes that the Einstein-Hilbert term $m^2 R$ is nonzero which makes it somewhat suspicious in the $m^2 \rightarrow 0$ limit. }

A final remark regarding the $m^2 \rightarrow 0$ limit: While one counts $2+5=7$ on-shell states in the massive theory (massless + massive gravitons), the massless Weyl theory has $6=2+2+2$ states (planewave + non-planewave gravitons +  photons) since a scalar should decouple due to the enhanced local scale symmetry~\cite{Stelle:1976gc}.  One might question the smoothness of the $m\to0$ limit since the propagator typically has $1/m^2$ poles; however, we have confirmed by explicit calculation that the amplitudes obtained using either the $m=0$ or $m\to0$ prescriptions are identical. This point is further discussed in appendix~\ref{sec:gaugefixing}.

\subsection{Minimal conformal supergravity}
\label{sec:minCSG}

In extending the discussion to conformal supergravity amplitudes, we seek a bosonic extension to the Weyl Lagrangian~(\ref{eq:weylLagrangian}) which originates from the SU(4)-singlet part of minimal ${\cal N}=4$ conformal supergravity.  Following Fradkin and Tseytlin's classification~\cite{Fradkin:1985am}, minimal conformal supergravity is defined as possessing both the global SU(1,1) symmetry acting on the scalars $\phi^\alpha$ and the local U(1) symmetry acting also on other fields.  Non-minimal theories, which we will discuss in section~\ref{sec:confgravity}, break both symmetries.  All dependence on $\phi^\alpha$ in the minimal theory should therefore be through the coset field $P^\mu$ and the U(1)-covariant derivative $\tilde{\nabla}_\mu$.  Given also that the Lagrangian should be of the four-derivative type and have local scale invariance, the only allowed SU(4)-singlet combinations are
\begin{align}\label{eq:confAnsatz}
(W_{\mu\nu\rho\sigma})^2, && 
\bP^\mu\tilde{\nabla}_\mu\tilde{\nabla}_\nu P^\nu+2(R_{\mu\nu}\!-\!\frac{1}{3}g_{\mu\nu}R)\bP^\mu P^\nu, &&
P^2\bP^2, && (P\!\cdot\!\bP)^2,
\end{align}
where total derivatives and the Gauss-Bonnet term are excluded.

As the U(1) symmetry decouples the scalars from tree-level graviton amplitudes, our discussion of the Weyl theory tells us that the pure-graviton superconformal amplitudes are given by $m^2$ times their Einstein supergravity counterparts.  For this to generalize to the full physical planewave spectrum of $\cN=4$ conformal supergravity --- the same $\cH^\pm$ multiplets as appeared in $\cN=4$ Einstein supergravity~(\ref{eq:neq4sgmultiplet}) --- we require the equations of motion for Einstein supergravity to imply those of the conformal theory.  In the SU(4)-singlet sector, the former are
\begin{align}\label{eq:EE}
R_{\mu\nu}-\frac{1}{2}g_{\mu\nu}R=T^{(2)}_{\mu\nu}\,, &&
\tilde{\nabla}_\mu P^\mu=0\,,
\end{align}
and the stress-energy tensor is
\begin{align}
T^{(2)}_{\mu\nu}=\frac{2}{e}\frac{\delta(eP\!\cdot\!\bP)}{\delta g^{\mu\nu}}
=2P_{(\mu}\bP_{\nu)}-g_{\mu\nu}P\!\cdot\!\bP\,.
\end{align}
This allows us to make an ansatz for the superconformal Lagrangian using the terms in~(\ref{eq:confAnsatz}), and fix coefficients by examining the resulting equations of motion.

Doing so, we find that the $\cN=4$ minimal conformal supergravity theory has the following SU(4)-singlet sector Lagrangian:
\begin{align}\label{eq:minConfAction} 
e^{-1}\cL&=-\frac{1}{4}(W_{\mu\nu\rho\sigma})^2+
\bP^\mu\tilde{\nabla}_\mu\tilde{\nabla}_\nu P^\nu+2(R_{\mu\nu}\!-\!\frac{1}{3}g_{\mu\nu}R)\bP^\mu P^\nu
-P^2\bP^2-\frac{1}{3}(P\!\cdot\!\bP)^2\nn\\
&\qquad+m^2\left(-\frac R2+P\cdot\bP\right),
\end{align}
where we have introduced a mass deformation by the Einstein supergravity terms~(\ref{eq:nmGravityP}) from the very beginning.  The resulting equations of motion are
\begin{subequations}\label{eq:CGeom}
\begin{align}
\label{eq:CGeoma}
  & B_{\mu\nu}+m^2\left(R_{\mu\nu}-\frac{1}{2}g_{\mu\nu}R\right)=T_{\mu\nu}^{(4)}\,,\\
\label{eq:CGeomb}
  & (\tilde{\nabla}^2+m^2-4P\!\cdot\!\bP)\tilde{\nabla}_\mu P^\mu=0\,.
\end{align}
\end{subequations}
The scalar equation is clearly implied by $\tilde{\nabla}_\mu P^\mu=0$. Showing that the graviton equation is also solved automatically is more involved.  The stress tensor is
\begin{align}\label{eq:TCG}
\begin{aligned}
T^{(4)}_{\mu\nu}&=m^2 T^{(2)}_{\mu\nu}-2(\tilde{\nabla}_{(\mu} P_{\nu)}\tilde{\nabla}_\rho\bP^\rho+\text{h.c.})+g_{\mu\nu}(\tilde{\nabla}_\rho P^\rho)(\tilde{\nabla}_\lambda \bP^\lambda)\\
&\quad-2\left(\nabla^\rho\nabla_\mu P_{(\rho}\bP_{\nu)}+\nabla^\rho\nabla_\nu P_{(\rho}\bP_{\mu)}\right)+2\square P_{(\mu}\bP_{\nu)}+\frac{4}{3}\nabla_\mu\nabla_\nu (P\!\cdot\!\bP)\\
&\quad-\frac{1}{3}g_{\mu\nu}\square(P\!\cdot\!\bP)+g_{\mu\nu}\left[2\nabla^\rho\nabla^\lambda P_{(\rho}\bP_{\lambda)}-\square(P\!\cdot\!\bP)\right]\\
&\quad+4g^{\rho\lambda}\left[(R_{\mu\rho}-2P_{(\mu}\bP_{\rho)})P_{(\lambda}\bP_{\nu)}+P_{(\mu}\bP_{\rho)}R_{\lambda\nu}\right]\\
&\quad-\frac{4}{3}\left[(R-2P\!\cdot\!\bP)P_{(\mu}\bP_{\nu)}+R_{\mu\nu}(P\!\cdot\!\bP)\right]\\
&\quad-2g_{\mu\nu}\left[(R_{\rho\lambda}-P_{(\rho}\bP_{\lambda)})P^{(\rho}\bP^{\lambda)}-\frac{P\!\cdot\!\bP}{3}(R-P\!\cdot\bP)\right]\,.
\end{aligned}
\end{align}
A helpful first step is to re-express the gravitational equation given in eq.~(\ref{eq:EE}) as $R_{\mu\nu}=2P_{(\mu}\bP_{\nu)}$ and $R=2P\!\cdot\!\bP$.  On support of these relations, $T^{(4)}_{\mu\nu}$ then reduces to
\begin{align}
T^{(4)}_{\mu\nu}=m^2T^{(2)}_{\mu\nu}+B_{\mu\nu}\,.
\end{align}
So the four-derivative gravitational equation of motions~(\ref{eq:CGeoma}) are indeed solved by the two-derivative counterparts.

As a final confirmation of this Lagrangian we compare it with the bosonic part of the action constructed by Ciceri and Sahoo for the same theory~\cite{Ciceri:2015qpa}.  This version of the action contains additional gauge fields associated with the symmetries of the conformal group, which allow the full conformal symmetry (including conformal boosts) to be realized covariantly.  As they do not correspond to physical states we remove them either by integration or gauge fixing --- for full details, see appendix~\ref{sec:cicerisahoo}.

The double copy structure of the amplitudes of minimal conformal gravity is inherited from the two-derivative theories, and follows the usual pattern for different degrees of supersymmetry,\footnote{Note that this double-copy identification is for the four-dimensional theories. For $D>4$ the Gauss-Bonnet term needs to be accounted for, which may alter the details.}
\begin{align}
(\cN=0,1,2,4\text{ min.CSG}) = \left(\text{min.}(DF)^2\right) \otimes (\cN=0,1,2,4\text{ SYM})\,.
\end{align}
Given that amplitudes involving planewave states vanish in both minimal $(DF)^2$ and minimal conformal (super)gravity the relation is trivially satisfied in this case. A less trivial confirmation is that this double copy also works when one external state is taken as a non-planewave mode. This works because the minimal $(DF)^2$ and minimal conformal supergravity amplitudes are proportional to the ordinary Yang-Mills and Einstein supergravity amplitudes, which obey the color-kinematics duality and the double copy, respectively. We leave further confirmation of the minimal-theory double copy to future work.

%**************************************************
\section{Non-minimal conformal supergravity}
\label{sec:confgravity}

We now proceed to the Lagrangian description of non-minimal conformal supergravities, and their associated scattering amplitudes.  The bosonic parts of the complete class of $\cN=4$ supersymmetric Lagrangians were recently constructed by Butter, Ciceri, de Wit and Sahoo~\cite{Butter:2016mtk}. The SU(4)-singlet part of their Lagrangian is 
\begin{align}\label{eq:nmConfAction}
e^{-1}\cL&=-\frac{\cF}{2}\bigg[
\frac{1}{2}(W^+_{\mu\nu\rho\sigma})^2\!-\!\bP^\mu\tilde{\nabla}_\mu\tilde{\nabla}_\nu P^\nu\!-
\!2(R_{\mu\nu}\!-\!\frac{1}{3}g_{\mu\nu}R)\bP^\mu P^\nu\!+\!P^2\bP^2\!+\!\frac{1}{3}(P\!\cdot\!\bP)^2\bigg]\nn\\
&\qquad+\text{h.c.}.
\end{align}
The different conformal supergravities are encoded in terms of the zeroth-degree homogeneous free function $\cF(\phi_\alpha)$. The SU(1,1)/U(1) coset field $P_\mu$ and U(1)-covariant derivative $\tilde{\nabla}_\mu$ were introduced in section~\ref{sec:cosetformulation}.   The (anti-)self-dual part of the Weyl tensor is defined as
\begin{align}
W^\pm_{\mu\nu\rho\sigma}&=
\frac{1}{2}\left(W_{\mu\nu\rho\sigma}\pm\tilde{W}_{\mu\nu\rho\sigma}\right), &&
\tilde{W}_{\mu\nu\rho\sigma}=
\frac{i}{2}e\epsilon_{\mu\nu\kappa\lambda}{W^{\kappa\lambda}}_{\rho\sigma},
\end{align}
where $W_{\mu\nu\rho\sigma}$ was given in eq.~(\ref{eq:weylTensor}).  In this case, rather than construct an ansatz, we obtained the above Lagrangian directly from ref.~\cite{Butter:2016mtk} by eliminating additional gauge fields associated with conformal symmetries --- for details, see appendix~\ref{sec:butter}.  

The scalar function $\cF$ is of primary interest to us, $\cF=1$ corresponding to the already-discussed minimal theory for which all planewave tree-level amplitudes vanish (for $m=0$).  For non-minimal  $\cN=4$ conformal supergravities the presence of $\cF$ breaks the SU(1,1)$\times$U(1) symmetry and allows for non-vanishing planewave amplitudes.  Using the parametrization of $\phi^\alpha$ in terms of $C$~(\ref{eq:cayleighParam}), we notice that any choice of $\cF$ that is solely dependent on $\bC=\phi_2/\phi_1$ (or, alternatively, dependent on $\bar{\tau}$) is a zeroth-degree homogeneous function as required.  Our main focus is
\begin{align}
\cF=i\bar{\tau},
\end{align}
for which we observe that the tree-level amplitudes match those of the Berkovits-Witten theory~(\ref{eq:BW}).\footnote{See also ref.~\cite{Tseytlin:2017qfd} where the conformal anomaly of the Berkovits-Witten theory was analyzed using the Lagrangian of ref.~\cite{Butter:2016mtk}.} However, before discussing this case in more detail we make some remarks about amplitudes for arbitrary $\cF$.

\subsection{Generic non-minimal amplitudes}

\def\jj{j}

Given that $\cF$ may be considered a function of only $\bC$, and that for sensible perturbations around $C=0$ it should also be analytic at the origin, we define it as a series expansion:
\begin{align}\label{eq:hExpansion}
\cF(\bC)=1-2\sum_{\jj=1}^\infty\frac{f_\jj}{\jj!}\bC^\jj.
\end{align}
Different theories are classified by the constants $f_\jj$.  As all terms inside the square brackets of the non-minimal Lagrangian~(\ref{eq:nmConfAction}) are at least quadratic in the fields, both the scalar and graviton propagators are unaffected by the $f_\jj$.  Therefore, the propagators and on-shell states of the different non-minimal conformal supergravities matches that of minimal theory, which, as we explained in section~\ref{sec:minDF2}, for physical planewave states matches that of ${\cal N}=4$ Einstein supergravity~(\ref{eq:neq4sgmultiplet}).  Our gauge fixing procedure and resulting propagators are described in full generality in appendix~\ref{sec:gaugefixing}.

The dependence on $\cF$ in the conformal supergravities is through the interactions.  When $\cF\neq1$ the equations of motion for Einstein supergravity~(\ref{eq:EE}) no longer imply those of the conformal theory. Therefore, tree-level conformal supergravity amplitudes with physical planewave states are generally not proportional to those of Einstein supergravity, hence they do not vanish.  Henceforth we focus only on the planewave states.  The nontrivial amplitudes from the non-minimal Lagrangian~(\ref{eq:nmConfAction}) can be calculated from Feynman diagrams.  However, obtaining tree-level amplitudes from solutions to the classical equations of motion is better.  Berends-Giele recursion~\cite{Berends:1987me} is our main tool for checking the amplitudes discussed in this and other sections.

At three points only the three-scalar and two-graviton-one-scalar vertices are modified with respect to the minimal theory.  The three-scalar amplitudes are, however, still required to be zero due to the three-point kinematics $p_i\cdot p_j=0$.  The only non-zero three-point amplitudes are therefore
\begin{align}
M_3(h^{--},h^{--},C)=if_1\braket{12}^4, && M_3(h^{++},h^{++},\bC)=if_1[12]^4\,.
\end{align}
For generic $f_1\neq0$ these amplitudes clearly violate the U(1) symmetry, as the scalar carries U(1) charge.  The amplitudes can be supersymmetrized to capture the full $\cN=4$ content; the MHV amplitude takes the form $M_3(\cH_1^-\cH_2^-\cH_3^-)=if_1\delta^8(Q)$.

Proceeding to $n$-point amplitudes, the Feynman vertices that are needed can only receive contributions from coefficients up to $f_{n-2}$, which itself first appears in the 2-graviton-$(n-2)$-scalar and $n$-scalar vertices.  Although the resulting amplitudes are more complicated, we can identify certain patterns by restricting the external states. For instance, in the MHV sector if all external states are restricted to the $\cH^-$ multiplet~(\ref{eq:neq4sgmultiplet}) then one can show that the amplitudes have no poles.  They may be expressed simply as
\begin{equation}
M_n(\cH_1^-\cH_2^-\cdots\cH_n^-)=iS_{n-2}\delta^8(Q),
\end{equation}
where $S_n(f_{\jj})$ is a constant function of $f_\jj$ for us to determine.

We have checked by explicit calculation that
\begin{subequations}\label{eq:allPlusAmps}
\begin{align}
M_3(h^{--},h^{--},C)&=if_1\braket{12}^4,\\
M_4(h^{--},h^{--},C,C)&=i(f_2+3f_1^2)\braket{12}^4,\\
M_5(h^{--},h^{--},C,C,C)&=i(f_3+10f_1f_2+15f_1^3)\braket{12}^4,\\
M_6(h^{--},h^{--},C,C,C,C)&=i(f_4+10f_2^2+15f_1f_3+105f_1^2f_2+105f_1^4)\braket{12}^4.
\end{align}
\end{subequations}
The numerical coefficients are clearly combinatoric factors --- for instance, the coefficient of $f_1f_2$ in the five-point amplitude is $10=5!/(2!\times3!)$, which counts the number of five-point diagrams involving a three- and a four-point vertex ($f_1$ is associated with the former and $f_2$ with the latter).  A recursive formula for $S_n$ generalizes this pattern to arbitrary multiplicity,
\begin{align}\label{eq:recursiveSN}
\frac{S_n}{(n+1)!}&=
\sum_{\jj=1}^nf_\jj\!\!\!\!\!\sum_{\substack{i_1,\ldots,i_{n-1}\\n=\jj+\sum_{k=1}^{n-1}i_kk}}\!\!\!
\frac{1}{(\jj \!+\!1\!-\!\sum_{k=1}^{n-1}i_k)!i_1!\cdots i_{n-1}!}
\bigg(\frac{S_1}{2!}\bigg)^{i_1}\!\!\!\!\cdots\bigg(\frac{S_{n-1}}{n!}\bigg)^{i_{n-1}}\!\!\!\!\!\!\!,
\end{align}
where the recursion starts from $S_0=1$.

The same pattern also applies when one external state belongs to $\cH^+$ in the MHV sector.  By checking $(n-1)$-scalar one-graviton amplitudes up to seven points and $(n-3)$-scalar three-graviton amplitudes up to five points we have confirmed that
\begin{align}\label{eq:onePlus}
M_n(\cH_1^+\cH_2^-\cdots\cH_n^-)=
-iS_{n-3}\delta^8(Q)\sum_{i=2}^n\frac{[1i]\braket{iq}^2}{\braket{1i}\braket{1q}^2}, \qquad n\geq4.
\end{align}
The amplitudes reduce to those of the all-minus sector~(\ref{eq:allPlusAmps}) in the soft limit $p_1\to0$, in which case the extra kinematic function has the interpretation of a soft factor.  This agrees with the BCFW-inspired soft factor of ref.~\cite{BoucherVeronneau:2011nm}; similar arguments using these soft factors have been used to calculate anomalous one-loop $\cN=4$ Einstein supergravity amplitudes~\cite{Carrasco:2013ypa,Bern:2017rjw}.

\subsection{The Berkovits-Witten theory}

We remarked earlier that $\cF=i\bar{\tau}$ gives rise to a theory whose MHV amplitudes match those of the Berkovits-Witten twistor string~(\ref{eq:BW}).  Using the generic all-minus amplitudes found in the previous subsection, this is easily confirmed.  Comparing the expansion of $\cF$~(\ref{eq:hExpansion}) with
\begin{align}
i\bar{\tau}=\frac{1+\bC}{1-\bC}=1+2\sum_{\jj=1}^\infty\bC^\jj,
\end{align}
we find that $f_\jj=-\jj !$.  In this case $S_n=(-1)^n$, so as expected the all-minus amplitudes are $M_n=i(-1)^n\delta^8(Q)$.  The one-plus amplitudes also agree: in the $n$-point expression~(\ref{eq:onePlus}) we recognize the helicity-dependent soft function as coming from the Berkovits-Witten formula~(\ref{eq:BW}).  Given that at $n$ points we uniquely fix $f_{n-2}$ using the all-minus sector, the identification $\cF=i\bar{\tau}$ is complete.

Using numerical Berends-Giele recursion we have explicitly checked this for a wide variety of scalar-graviton amplitudes up to $n=7$, including $M_6(C,C,C,\bar{C},\bar{C},\bar{C})$.  While this NMHV amplitude is not predicted by the Berkovits-Witten formula, we have compared it with NMHV amplitudes coming from the double copy $(DF)^2\otimes {\rm YM}$. This provides nontrivial evidence for the validity of this double-copy construction for a wider class of amplitudes than those considered by Berkovits and Witten.\footnote{{This numeric check was made against both the $(DF)^2$ Lagrangian of ref.~\cite{Johansson:2017srf} and the $(DF)^2$ CHY integrand of ref.~\cite{Azevedo:2017lkz}. Ref.~\cite{Farrow:2018cqi} provides a toolkit for the evaluation of such amplitudes in the CHY framework.}}

\subsection{The mass-deformed theory}\label{sec:massdef}

When studying minimal conformal supergravity we found it helpful to deform the theory by a massive component proportional to the Einstein supergravity action, thus realizing the ghost-like particles as massive states by breaking scale symmetry.  In non-minimal theories we can do the same,
\begin{align}\label{eq:csgNMmassdef}
e^{-1}\cL&=-\frac{\cF}{2}\bigg[
\frac{1}{2}(W^+_{\mu\nu\rho\sigma})^2\!-\!\bP^\mu\tilde{\nabla}_\mu\tilde{\nabla}_\nu P^\nu\!-
\!2(R_{\mu\nu}\!-\!\frac{1}{3}g_{\mu\nu}R)\bP^\mu P^\nu\!+\!P^2\bP^2\!+\!\frac{1}{3}(P\!\cdot\!\bP)^2\bigg]\nn\\
&\qquad+\text{h.c.}+m^2\left(-\frac{R}{2}+P\cdot\bP\right).
\end{align}
With $\cF=i\bar{\tau}$ the amplitudes of this theory are expected to arise from the double copy between mass-deformed $(DF)^2$ theory (\ref{eq:df2MassDef}) and $\cN=4$ SYM.  They smoothly interpolate between Berkovits-Witten conformal supergravity as $m\to0$ and $\cN=4$ Einstein supergravity as $m\to\infty$~\cite{Johansson:2017srf}.  For
instance, the following four-point amplitudes depend non-trivially on the mass:
\begin{subequations}\label{eq:4ptmassdef}
\begin{align}
M_4(\cH_1^-\cH_2^-\cH_3^-\cH_4^-)&=i\delta^8(Q)\frac{stu+2m^6}{(s-m^2)(t-m^2)(u-m^2)},\\
M_4(\cH_1^+\cH_2^+\cH_3^-\cH_4^-)&=i\delta^8(Q)\frac{[12]^4}{st}\left(\frac{t}{s-m^2}+\frac{m^2}{u}\right)\,,
\end{align}
\end{subequations}
while the remaining four-point amplitude $M_4(\cH_1^+\cH_2^-\cH_3^-\cH_4^-)$ is independent of mass.  All three reproduce the Berkovits-Witten amplitudes~(\ref{eq:BW}) when $m=0$; the ghostlike internal states are exposed as poles of the form $(p^2-m^2)^{-1}$, including in the all-minus sector where previously the poles could not be resolved. The $\cN=4$ Einstein supergravity amplitudes are obtained after dividing by $m^2$ (in order to get amplitudes of correct dimension) and sending $m\to\infty$ --- only the $M_4(\cH_1^+\cH_2^+\cH_3^-\cH_4^-)$ amplitude contributes in this limit. 

We have checked these three amplitudes by explicit calculation, using both the double copy and the conformal supergravity Lagrangian~(\ref{eq:csgNMmassdef}).  We have also numerically cross-checked the corresponding five-point amplitudes with those arising from the double copy.  Again, the details of the gauge-fixing procedure for performing these checks starting from the Lagrangian~(\ref{eq:csgNMmassdef}) may be found in appendix~\ref{sec:gaugefixing}.

\subsection{Coupling to vector multiplets}
\label{sec:weylSYM}

Another important deformation to the $(DF)^2$ theory considered in ref.~\cite{Johansson:2017srf} was with terms containing bi-adjoint scalars $\phi^{aA}$~(\ref{eq:DF2YMphi3}), for which a coupling of the gravitational theory to non-abelian vector multiplets $\cV$~(\ref{eq:neq4multiplet}) was anticipated after the double copy.  Knowing already how these non-abelian vectors couple to Einstein supergravity, we anticipate their coupling to the conformal supergravity with $\cF=i\bar{\tau}$ as
\begin{align}
\begin{aligned}\label{eq:csgwVectors}
e^{-1}\cL&=-\frac{i\bar{\tau}}{2}\bigg[
\frac{1}{2}(W^+_{\mu\nu\rho\sigma})^2+\frac{1}{2}(F^{+,A}_{\mu\nu})^2-\bP^\mu\tilde{\nabla}_\mu\tilde{\nabla}_\nu P^\nu+P^2\bP^2+
\frac{1}{3}(P\!\cdot\!\bP)^2\\
&\qquad-2(R_{\mu\nu}\!-\!\frac{1}{3}g_{\mu\nu}R)\bP^\mu P^\nu\bigg]+\text{h.c.}
+m^2\left(-\frac{R}{2}+P\cdot\bP\right)\,,
\end{aligned}
\end{align}
where as usual only the SU(4)-singlet part is given, and the self-dual part of the YM field strength $F^{+,A}_{\mu\nu}$ was introduced in eq.~(\ref{eq:selfdualCurvature}). For completeness we have also included the mass deformation. 

First we consider vectors in the undeformed theory, $m=0$, which is the full-fledged Weyl-YM theory coming from Witten's twistor string. The Berkovits-Witten MHV formula~(\ref{eq:BW}) generalizes to the single-trace sector as
\begin{align}\label{eq:genBW}
M_n(\cH_1^+\cdots\cH_k^+\cH_{k+1}^-\cdots\cH_{r-1}^-\cV_{r}\cdots\cV_n)=
i\frac{(-1)^{r-1}\delta^8(Q)}{\braket{r,r\!+\!1}\cdots\braket{nr}}\prod_{i=1}^k
\sum_{\substack{j=1 \\ j\neq i}}^n\frac{[ij]\langle jq\rangle^2}{\langle ij\rangle\langle iq\rangle^2}.
\end{align}
By explicit calculation, again using both the double copy and starting from the Lagrangian~(\ref{eq:csgwVectors}), we have confirmed this for a wide variety of MHV amplitudes involving gravitons, scalars and gluons up to seven points.  There is a clear similarity between these amplitudes and the supersymmetric Einstein-Yang-Mills (EYM) amplitudes given in eq.~(\ref{eq:mhvEYMSG}); interestingly though, here the formula holds regardless of how many $\cH^-$ multiplets are scattered.  Also by analogy to the EYM version~(\ref{eq:mhvEYMSGDoubleTrace}), the double-trace MHV amplitudes are
\begin{align}\label{eq:confDoubleTrace}
M_n(\cV_1\cdots \cV_{r-1}|\cV_r\cdots\cV_n)=
i\frac{\delta^8(Q)}{\braket{12}\cdots\braket{r\!-\!1,1}\braket{r,r\!+\!1}\cdots\braket{nr}}.
\end{align}
The only difference with EYM is the lack of an overall $-p_{r,n}^2$ factor, i.e. the momentum exchanged between the two color traces.  This is easily understood by realizing that the graviton propagator is now $-1/p^2$ times the two-derivative propagator (see appendix~\ref{sec:gaugefixing}).

Finally we consider vector multiplets in the $m\neq0$ theory.   By explicit calculation, we confirm a smooth interpolation between the undeformed superconformal amplitudes as $m\to0$ and $\cN=4$ EYM supergravity amplitudes as $m\to\infty$ (after multiplication by appropriate power of $m^2$).  At four points, in addition to the two given in section~\ref{sec:massdef}, the amplitudes that depend on $m$ are 
\begin{subequations}
\begin{align}
M_4(\cH_1^-\cH_2^-\cV_3\cV_4)&=-i\frac{s}{s-m^2}\frac{\delta^8(Q)}{\braket{34}^2},\\
M_4(\cV_1\cV_2|\cV_3\cV_4)&=i\frac{s}{s-m^2}\frac{\delta^8(Q)}{\braket{12}^2\braket{34}^2}.
\end{align}
\end{subequations}
The massive pole is in both cases apparent, and can be thought as arising from swapping an $s^{-1}$ pole with $(s-m^2)^{-1}$.  The latter generalizes at $n$ points MHV to
\begin{align}
M_n(\cV_1\cdots \cV_{r-1}|\cV_r\cdots\cV_n)=i \frac{p^2_{r,n}}{p^2_{r,n}-m^2}
\frac{\delta^8(Q)}{\braket{12}\cdots\braket{r\!-\!1,1}\braket{r,r\!+\!1}\cdots\braket{nr}},
\end{align}
which we have checked by explicit calculation up to $n=7$.

%**************************************************
\section{Conclusions}
\label{sec:conclusions}

In this paper we have confirmed that a wide variety of conformal supergravity tree amplitudes have compatible Lagrangian and double-copy origins.  In particular, we have identified a non-minimal $\cN=4$ conformal supergravity Lagrangian whose physical planewave amplitudes are those of the Berkovits-Witten theory~\cite{Berkovits:2004jj}.  This theory turns out to be a simple case of the infinite class of non-minimal Lagrangians derived by Butter, de Wit, Ciceri and Sahoo~\cite{Butter:2016mtk}. The tree amplitudes are predicted by a double copy between $\cN=4$ super-Yang-Mills and a $(DF)^2$ theory, where the latter gauge theory was introduced by Nohle and one of the present authors in ref.~\cite{Johansson:2017srf}.  We have checked this double-copy construction by comparing to explicit calculations of conformal supergravity amplitudes from the Lagrangian, including six-point NMHV amplitudes which were not given in the original work~\cite{Berkovits:2004jj}.

Conformal supergravity theories have four-derivative kinetic terms, and as such they exhibit ghost-like states that are connected to non-unitary behavior. More specifically, four-derivative theories generally include both regular planewave modes and unusual non-planewave modes that exhibit linear growth  --- the two are sometimes described as forming a dipole.  Whether the non-planewave modes are acceptable asymptotic states for the S-matrix is an interesting question~\cite{Adamo:2018srx, Farrow:2018yqf}; our partial analysis seems to suggest that these states can be considered even if amplitudes for such states may pick up a dependence on a constant auxiliary vector $\alpha^\mu$.  As the four-derivative off-shell propagators are of the form $1/p^4$, the factorization properties of amplitudes when internal lines go on-shell are somewhat delicate.  We have showed that the leading $1/p^4$ poles are uniquely associated with intermediate planewave exchange, whereas subleading poles $1/p^2$ generically arise from internal non-planewave states propagating into planewave states. Thus the planewave and non-planewave states are not orthogonal.

The details of the states and propagator poles becomes clearer by introducing a mass deformation, which we did for the various four-derivative theories considered in this paper.  This allows for a unique separation of the degrees of freedom into massless and massive states, and amplitudes can be obtained using the standard LSZ procedure. In the $m\rightarrow 0$ limit details of the massless theories can be more easily inferred. Non-planewave modes emerge as an infinitesimal difference between the massless and massive modes, which is a prescription that generalizes to the amplitudes of such states. We use the prescription to compute tree amplitudes that have one external non-planewave mode, giving us a handle on certain minimal theories where the planewave modes give a vanishing tree-level S-matrix.

We found it useful to analyze the tree-level amplitudes of different four-derivative theories by using perturbative solutions to the classical equations of motion, where external states are specified as boundary conditions.  An analogous setup was used by Maldacena for pure Weyl conformal gravity in AdS space~\cite{Maldacena:2011mk}. The equations of motion are automatically solved by the vacuum Einstein equations $R_{\mu\nu}=0$, which implies vanishing of the conformal gravity planewave tree amplitudes in the flat-space limit~\cite{Adamo:2013tja,Adamo:2016ple,Beccaria:2016syk}.  By considering mass-deformed theories in flat space, we extended this argument to \emph{any} four-derivative theory whose equations of motion are implied by those of a two-derivative theory.  Such theories we refer to as being minimal, since the minimal conformal supergravities ($\cN=4$ and truncations thereof) fall into this class.  Knowing that the equations of motion for $\cN=4$ Einstein supergravity should imply those of minimal $\cN=4$ conformal supergravity, we derived the Lagrangian of the  latter theory from first principles.  Other minimal four-derivative theories that we considered include a scalar toy model studied in section~\ref{sec:toymodel} and the minimal $(DF)^2$, both of which have vanishing tree-level S-matrices for planewave states.

For the minimal theories we showed that amplitudes with a single non-planewave state are non-vanishing, and moreover equal (up to a sign) to amplitudes in the corresponding two-derivative theories.  Using the properties of amplitudes of the minimal theories, we have obtained sufficiently non-trivial evidence to support identifying minimal $\cN=4$ conformal supergravity with a new double-copy construction involving minimal $(DF)^2$ and $\cN=4$ super-Yang-Mills theory.  Further work is required to confirm that the double-copy construction continues to hold for tree amplitudes with more non-planewave modes, as well as the other states present in $\cN=4$ conformal supergravity in the form of gravitino multiplets~\cite{Berkovits:2004jj,Ciceri:2015qpa}. These questions we leave for future work.

Non-minimal four-derivative theories differ from minimal theories only by their interactions.  In $\cN=4$ conformal supergravity, these new interactions arise from a function $\cF$ of the complex scalars that multiplies the self-dual (Weyl)$^2$ term. For $\cF\neq1$ the SU(1,1)/U(1) symmetry enjoyed by the minimal theory is broken, and this allows for non-zero planewave amplitudes.  The choice $\cF=i\bar{\tau}$, where $\tau=\chi+ie^{-\varphi}$ is the usual complexified axion-dilaton field, gives the Berkovits-Witten version of the theory.  The appearance of $i\bar{\tau}$ in front of the self-dual (Weyl)$^2$ term is analogous to the vector field strength's coupling Einstein supergravity, where SU(1,1)/U(1) symmetry is also broken.  After including such terms for vector multiplets in the Berkovits-Witten conformal supergravity Lagrangian, we calculated the corresponding amplitudes and showed that they match the double-copy construction for Weyl-YM theories given in ref.~\cite{Johansson:2017srf}. 

We have also calculated certain amplitudes for generic choices of $\cF$, and showed that they take on a very simple form in the sector where all graviton multiplets are chiral. It would be interesting to explore whether a double-copy construction exists for generic choices of $\cF$. The question can be rephrased as whether it is possible to find  further variants of the $(DF)^2$ theory, by modifying interactions and field content, while at the same time preserving the color-kinematics duality of the theory.  Lifting the $(DF)^2$ theory to $D=10$ dimensions suggest that there are no obvious deformations in $D=10$, since the corresponding $\cN=1$ conformal supergravity theory has been argued to be unique~\cite{deRoo:1991at} (and should presumably be identified with the dimensionally-oxidized Berkovits-Witten theory). Similar uniqueness results can be inferred from the string theory double copies of ref.~\cite{Azevedo:2018dgo} that also involve the $(DF)^2$ theory in ten dimensions.  However, in lower dimensions --- such as $D=4$ or $D=6$ --- there might be room for further deformations of the $(DF)^2$ theory.  Alternatively, it is possible that the generic non-minimal $\cN=4$ conformal supergravity theories come from constructions of the type $(\cN=2)\otimes(\cN=2)$, where the unknown gauge theories are not constrained by maximal supersymmetry.

A natural extension to this work is to compute loop amplitudes in conformal supergravity, in order to better study ultraviolet properties, as well as issues with unitarity. The issue of unitarity will be more pressing, as ghosts  and other states now necessarily appear in the loops. This includes Faddeev-Popov ghosts that should be added to the $(DF)^2$ theory if we wish to use the double copy construction for off-shell loop momenta, without resorting to unitarity techniques. Fradkin and Tseytlin~\cite{Tseytlin:2017qfd,Fradkin:1983tg,Fradkin:1985am,Romer:1985yg} have shown that $\cN=4$ conformal supergravity, of either minimal or non-minimal type, have no conformal anomalies given that four vector multiplets are added to the spectrum. This suggest that the loop-level amplitudes of these gravity theories are ultraviolet finite to all orders, which would be interesting to confirm by explicit calculation.

\begin{acknowledgments}
The authors would like thank Tim Adamo, Thales Azevedo, Marco Chiodaroli, Arthur Lipstein, Oliver Schlotterer and Arkady Tseytlin for useful discussions.  This research is supported by the Swedish Research Council under grant 621-2014-5722, the Knut and Alice Wallenberg Foundation under grant KAW 2013.0235, and the Ragnar S\"{o}derberg Foundation under grant S1/16. 
\end{acknowledgments}

\appendix

%**************************************************
\section{Gauge fixing and the graviton propagator}
\label{sec:gaugefixing}

The quadratic graviton terms in the Lagrangians of all conformal (super)gravities are the same, so this analysis is applicable to both minimal and non-minimal theories. The quadratic terms are
\begin{align}
\begin{aligned}\label{eq:quadraticL}
  \cL&=-\frac{1}{2}h^{\mu\nu}\square(\square+m^2)h_{\mu\nu}+\frac{1}{6}h\square(\square+3m^2)h-\frac{1}{3}h\partial_{\mu}\partial_{\nu}(\square+3m^2)h^{\mu\nu}\\
  &\qquad +h^{\mu\nu}\partial_{\nu}\partial_{\rho}(\square+m^2)h^{\rho}_{~\mu}-\frac{1}{3}h^{\mu\nu}\partial_{\mu}\partial_{\nu}\partial_{\rho}\partial_{\lambda}h^{\rho\lambda}+\cO(h^3)\,,
\end{aligned}
\end{align}
where $h={h^\mu}_\mu$ (indices are raised and lowered with the Minkowski metric $\eta_{\mu\nu}$).  Diffeomorphism symmetry  $h_{\mu\nu}\to h_{\mu\nu}+2\partial_{(\mu}\xi_{\nu)}$ is respected for all $m$; however, dilatation symmetry $h_{\mu\nu}\to h_{\mu\nu}-2\lambda_D\eta_{\mu\nu}$ of the quadratic terms is respected only when $m=0$.  We therefore perform gauge fixing in these two cases separately --- for $m=0$ we use gauge-fixing terms
\begin{align}
\mathcal{L}_\text{GF}=
(\partial_\nu h^{\mu\nu}-\frac{1}{2}\partial^\mu h)\square(\partial^\rho h_{\mu\rho}-\frac{1}{2}\partial_\mu h)
+\frac{1}{3}(\square h-\partial^\mu\partial^\nu h_{\mu\nu})^2\,,
\end{align}
where the first fixes diffeomorphism symmetry and the second dilatation symmetry. The  $m=0$ graviton propagator is then
\begin{align}\label{eq:masslessprop}
\begin{tikzpicture}
  [baseline={([yshift=-.5ex]current bounding box.center)},thick,inner sep=0pt,minimum size=0pt,>=stealth,scale=0.5]
  \node (1) at (0,0) {};
  \node (2) at (4,0) {};
  \draw[boson] (1) node[left=0.1] {$h_{\mu\nu}$} to node[below=0.2] {$p$} (2) node[right=0.1] {$h_{\rho\sigma}$} {};
\end{tikzpicture}\,\,
=-\frac{i}{2}\frac{\eta_{\mu\rho}\eta_{\nu\sigma}+\eta_{\mu\sigma}\eta_{\nu\rho}-\eta_{\mu\nu}\eta_{\rho\sigma}}{p^4}\,.
\end{align}
When $m\neq0$ we instead use
\begin{align}
\cL_\text{GF}=(\partial_\nu h^{\mu\nu})(\square+m^2)(\partial^\rho h_{\mu\rho})
\end{align}
to fix diffeomorphism symmetry.  Although the resulting propagator is rather non-trivial, it can be reduced using the observation that any numerator terms containing projectors with explicit $p_\mu$ are irrelevant.  They cancel out in connected correlators as $p_\mu J^{\mu\nu}=0$, where $J^{\mu\nu}$ is the Berends-Giele current.  It is therefore sensible to use
\begin{equation}\label{eq:propdeformed}
  \begin{tikzpicture}
  [baseline={([yshift=-.5ex]current bounding box.center)},thick,inner sep=0pt,minimum size=0pt,>=stealth,scale=0.5]
  \node (1) at (0,0) {};
  \node (2) at (4,0) {};
  \draw[boson] (1) node[left=0.1] {$h_{\mu\nu}$} to node[below=0.2] {$p$} (2) node[right=0.1] {$h_{\rho\sigma}$} {};
\end{tikzpicture}=\!\frac{i}{2m^2}\!\left(\frac{\eta_{\mu\rho}\eta_{\nu\lambda}\!+\!\eta_{\mu\lambda}\eta_{\nu\rho}\!-\!\eta_{\mu\nu}\eta_{\rho\lambda}}{p^2}\!-\!\frac{\eta_{\mu\rho}\eta_{\nu\lambda}\!+\!\eta_{\mu\lambda}\eta_{\nu\rho}\!-\!\frac{2}{3}\eta_{\mu\nu}\eta_{\rho\lambda}}{p^2-m^2}\right)\!.
\end{equation}
While this resembles the partial-fractioned propagators we have seen earlier, it cannot be brought together into a factorized form.  This reflects the fact that when $m\neq0$ there is an extra on-shell massive graviton state, which decouples when $m=0$ due to the introduction of dilatation symmetry.

When computing conformal supergravity amplitudes involving only physical planewave modes, a useful check on our results has been to compare the $m=0$ gauge fixing with $m\neq0$ in the $m\to0$ limit, and in all cases we have found agreement.  One can see this from the propagators above: when $m=0$ the trace part of the Berends-Giele current ${J^\mu}_\mu=0$, so trace parts of the propagators are irrelevant.  In this case the massive propagator can be brought into a form equivalent to the massless propagator~(\ref{eq:masslessprop}).

%**************************************************
\section{Covariant description of conformal supergravities}
\label{sec:cicerisahoo}

In this appendix we demonstrate the equivalence between the four-dimensional conformal supergravity Lagrangians used in this paper and those already given in the supergravity literature~\cite{Ciceri:2015qpa,Butter:2016mtk}.  The key distinction is that, in addition to the vierbein ${e_\mu}^a$ and the spin connection ${\omega_\mu}^{ab}$ --- where $a,b,\ldots$ are tangent-space indices --- these alternative superconformal constructions also contain the gauge fields $b_\mu$ and ${f_\mu}^a$ associated with dilatations and conformal boosts respectively.  To make contact with our results we therefore eliminate these fields: the former by gauge fixing special conformal symmetry, and the latter by explicitly solving for it.\footnote{Except for our use of the mostly-minus metric $\eta_{ab}=\text{diag}(+,-,-,-)$, our conventions match those of Freedman and Van Proeyen~\cite{Freedman:2012zz}.  Chapter 15 of this book provides an excellent introduction to the superconformal construction of gravitational theories.}

\subsection{The minimal theory}

The bosonic terms in the minimal four-dimensional $\cN=4$ conformal supergravity Lagrangian are given by Ciceri and Sahoo~\cite{Butter:2016mtk}.  The SU(4)-singlet part is
\begin{align}\label{eq:minconfaction}
e^{-1}\cL&=-\frac{1}{2}\bigg[
\frac{1}{2}R(M)^{abcd}R(M)^+_{abcd}+P^2\bP^2+\frac{1}{3}(P\!\cdot\!\bP)^2\nn\\
&\qquad-2\bP^a[\cD_a\cD^bP_b+\cD^2P_a]-2\cD^aP^b\cD_a\bP_b-\cD^aP_a\cD^b\bP_b
\bigg]+\text{h.c.}, 
\end{align}
where $R(M)^+_{abcd}$ is the self-dual supercovariant curvature.  Derivatives $\cD_a$ (containing the new field content) are covariant under both conformal and local U(1) transformations.  The local SU(1,1)/U(1) coset fields $P_a$ and $\bP_a$ are defined using these derivatives:
\begin{align}
P_a=\phi^\alpha\eps_{\alpha\beta}\cD_a\phi^\beta, && \bP_a=-\phi_{\alpha}\eps^{\alpha\beta}\cD_a\phi_{\beta},
\end{align}
The U(1) gauge field $a_\mu$ solves $\phi^\alpha\cD_a\phi_\alpha=0$ (ignoring the fermionic contribution).

In addition to showing equivalence with the minimal Lagrangian~(\ref{eq:nmConfAction}) given in the main body of this paper, it is also instructive to demonstrate this Lagrangian's invariance under conformal and U(1) transformations.  As discussed in section~\ref{sec:cosetformulation}, the scalars $\phi^\alpha$ transform under only U(1) with weight 1.  The gauge fields transform under both kinds of transformations as~\cite{Freedman:2012zz}
\begin{subequations}
\begin{align}
\delta{e_\mu}^a&=-\lambda_D{e_\mu}^a,\\
\delta{\omega_\mu}^{ab}&=-4{\lambda_K}^{[a}{e_\mu}^{b]},\\
\delta b_\mu&=\partial_\mu\lambda_D+2 {\lambda_K}^ae_{\mu a},\\
\delta\!{f_\mu}^a&=\partial_\mu{\lambda_K}^a-b_\mu{\lambda_K}^a+{\omega_\mu}^{ab}\lambda_{K,b}+\lambda_D{f_\mu}^a,\\
\delta a_\mu&=\partial_\mu\lambda_A,
\end{align}
\end{subequations}
where $\lambda_D(x)$ and ${\lambda_K}^a(x)$ are the local gauge parameters associated with dilatations and conformal boosts respectively.

Conformal-covariant derivatives $\cD_a={e_a}^\mu\cD_\mu$ are assembled using the general principle that, for any argument $X$,
\begin{align}
\cD_\mu X=D_\mu X-\delta X|_{\lambda_D\to b_\mu,\,{\lambda_K}^a\to{f_\mu}^a,\,\lambda_A\to a_\mu},
\end{align}
where $D_\mu$ is the ordinary (non-conformal) covariant derivative containing the spin connection ${\omega_\mu}^{ab}$ (not to be confused with the gauge-covariant derivative~(\ref{eq:fieldstrengthdef})).  For instance, $\cD_\mu\phi^\alpha=(\partial_\mu-ia_\mu)\phi^\alpha$ matches $\tilde{\nabla}_\mu\phi^\alpha$ as given in section~\ref{sec:cosetformulation}; similarly,
\begin{subequations}\label{eq:covDs}
\begin{align}
\cD_aP^b&=D_aP^b-{e_a}^\mu(b_\mu+2ia_\mu)P^b,\\
\cD_a\bP^b&=D_a\bP^b-{e_a}^\mu(b_\mu-2ia_\mu)\bP^b,\\
\cD_a\cD_bP^c&=D_a\cD_bP^c-{e_a}^\mu(2(b_\mu+ia_\mu)\cD_bP^c+2\delta_b^cf_{\mu d}P^d-4\eta_{bd}{f_\mu}^{(c}P^{d)})~\label{eq:examplecovderiv}
\end{align}
% \begin{align}
% \cD_aP^b&=D_aP^b-b_aP^b,\\
% \cD_a\bP^b&=D_a\bP^b-b_a\bP^b,\\
% \cD_a\cD_bP^c&=D_a\cD_bP^c-{e_a}^\mu(2b_\mu\cD_bP^c+2\delta_b^cf_{\mu d}P^d-4\eta_{bd}{f_\mu}^{(c}P^{d)})~\label{eq:examplecovderiv}
% \end{align}
\end{subequations}
give us all required terms in the action~(\ref{eq:minconfaction}).

Dilatation symmetry of the Lagrangian~(\ref{eq:minconfaction}) follows trivially from the covariance of all terms --- $\delta_De^{-1}=4\lambda_De^{-1}$ cancels the overall scaling.  Verifying special conformal symmetry is a little harder --- some helpful intermediate results are
\begin{subequations}
\begin{align}
\delta_K(\cD_aP^b)&=2\delta_a^b{\lambda_K}^cP_c-4\eta_{ac}{\lambda_K}^{(b}P^{c)},\\
\delta_K(\cD_a\bP^b)&=2\delta_a^b{\lambda_K}^c\bP_c-4\eta_{ac}{\lambda_K}^{(b}\bP^{c)},\\
\delta_K(\cD_a\cD_bP^c)&=
2\eta_{ab}{\lambda_K}^d\cD_dP^c-8\lambda_{K,(a}\cD_{b)}P^c-
4{\lambda_K}^c\cD_{(a}P_{b)}+4\lambda_{K,d}\delta^c_{(a}\cD_{b)}P^d.
\end{align}
\end{subequations}
The first of these has already been used to write down the covariant derivative~(\ref{eq:examplecovderiv}).  One ultimately finds that
\begin{align}
\delta_K(e^{-1}\cL)=-8\lambda_{K,a}\bP_b\cD^{[a}P^{b]}+\text{h.c.},
\end{align}
which vanishes on support of the Maurer-Cartan equations associated with the SU(1,1)/U(1) coset space~\cite{Ciceri:2015qpa,Butter:2016mtk}.

To reproduce the minimal Lagrangian given in the main body of this paper~(\ref{eq:minConfAction}) (not including its mass deformation) we eliminate the additional field content.  First, $\phi^\alpha\cD_a\phi_\alpha=0$ is solved to give $a_\mu=i\phi^\alpha\partial_\mu\phi_\alpha$.  Special conformal transformations are gauge fixed by setting $b_\mu=0$, giving $\lambda_{K,a}=-\frac{1}{2}{e_a}^\mu\partial_\mu\lambda_D$.  Then we use the following constraints on superconformal curvatures~\cite{Ciceri:2015qpa}:
\begin{subequations}
\begin{align}
{R(P)_{\mu\nu}}^a&=0,\\
{R(M)_{\mu\nu}}^{ab}{e^\nu}_b&=0.
\end{align}
\end{subequations}
The first of these identifies $\omega=\mathring{\omega}[e]$, which is the torsion-free spin connection.  The second is solved for ${f_\mu}^a$: substituting
\begin{align}
{R(M)_{\mu\nu}}^{ab}={R_{\mu\nu}}^{ab}+8{f_{[\mu}}^{[a}{e_{\nu]}}^{b]},
\end{align}
it is straightforward to show that
\begin{align}
{f_\mu}^a=-\frac{1}{4}\left({R_\mu}^a-\frac{1}{6}{e_\mu}^aR\right).
\end{align}
It follows that $W_{\mu\nu\rho\sigma}={R(M)_{\mu\nu}}^{ab}e_{\rho,a}e_{\sigma,b}$ and therefore $R(M)^{abcd}R(M)^+_{abcd}=(W^+_{\mu\nu\rho\sigma})^2$.  When this expression for ${f_\mu}^a$ is substituted into the Lagrangian~(\ref{eq:minconfaction}), together with the covariant derivatives~(\ref{eq:covDs}) and the gauge fixing $b_\mu=0$, the Lagrangian~(\ref{eq:nmConfAction}) is reproduced up to topological terms.

\subsection{The non-minimal theory}
\label{sec:butter}

The bosonic part of the complete Lagrangian for all $\cN=4$ conformal supergravities has been constructed by Butter, Ciceri, de Wit and Sahoo~\cite{Butter:2016mtk}.  Its SU(4)-singlet part is
\begin{align}\label{eq:confAction}
\begin{aligned}
e^{-1}\cL&=-\frac{\cF}{2}\bigg[
\frac{1}{2}R(M)^{abcd}R(M)^+_{abcd}-\bP^a\cD_a\cD_bP^b+
P^2\bP^2+\frac{1}{3}(P\!\cdot\!\bP)^2\\
&\qquad+4{e_a}^\mu{f_\mu}^c\eta_{cb}[P^a\bar{P}^b-P^d\bP_d\eta^{ab}]\bigg]
+\text{h.c.},
\end{aligned}
\end{align}
where $\cF(\phi_\alpha)$ is the zeroth-degree homogeneous function introduced in the main text.  When $\cF=1$, by dropping total-derivative terms one may re-express this result as Ciceri and Sahoo's minimal Lagrangian~(\ref{eq:minconfaction}).  By a completely analogous procedure to that used for the minimal Lagrangian, we have eliminated $b_\mu$ and ${f_\mu}^a$ to obtain the version~(\ref{eq:nmConfAction}) given in the main text.

\bibliographystyle{JHEP}
\bibliography{references}

\end{document}